\documentclass[preprint]{JHEP3} 

\usepackage{graphicx}
\usepackage{psfrag}

\usepackage{epsfig,multicol,bbm}

\title{
The Chiral Condensate in a Finite Volume
}
\author{
Poul ~H.~Damgaard$^a$,
Hidenori~Fukaya$^{a,b}$,\\
\it $^a$ The Niels Bohr International Academy, The Niels Bohr Institute,
Blegdamsvej 17 DK-2100 Copenhagen {\O}
Denmark \\
\it $^b$ High Energy Accelerator Research Organization (KEK),
  Tsukuba 305-0801, Japan\\
Email: 
\email{phdamg@nbi.dk},
\email{hfukaya@nbi.dk},
}

\abstract{
Chiral perturbation theory at finite four-volume $V(=L^3 T)$
is reconsidered with a view towards finding a computational
scheme that can deal with any value of $M_{\pi}L$, where
$M_{\pi}$ is a generic Nambu-Goldstone mass. The momentum zero modes that cause
the usual $p$-expansion to fail in the chiral limit
are treated separately, and partly integrated out to all
orders. In this way the theory remains infrared
finite in the perturbative expansion, and the chiral
limit can be considered at finite volume.
We illustrate the technique
by computing the quark condensate in a finite volume, 
smoothly connecting standard results in the $p$-regime for
larger masses with those of the
$\epsilon$-regime for smaller masses.
From the partially quenched theory we also
obtain the spectral density of the Dirac operator, a
smooth function from the microscopic region to the bulk region
of the $p$-regime.
}

\begin{document}

\section{Introduction}
\label{sec:intro}
\setcounter{equation}{0}

A number of studies have been devoted to
the finite volume effects in low-energy QCD, 
or chiral perturbation theory (ChPT) 
\cite{Gasser:1987ah}-\cite{Leutwyler:1992yt}.
The resulting finite-size scaling theory in ChPT has 
broad theoretical interest as it describes 
the critical behavior of dynamical symmetry breaking,
and as such has corresponding applications in statistical
physics as well. In lattice gauge theory it is certainly of great 
practical value to have analytical predictions for finite
volumes available, as they can help in eliminating 
uncertainties due to the finite sizes used in numerical
simulations \cite{BesselfiniteV}.  

To investigate the finite-size behavior of ChPT,
essentially two perturbative approaches have been proposed so far
\cite{Gasser:1987ah}.
One is the $p$-expansion, which has just the same
form as the perturbative series in an infinite volume,
only replacing momentum integrals by
the discrete sums over momentum due to the quantization
in units of $1/L$ (where $L$ is the linear extent).
If we denote by $M_{\pi}$ the mass of a generic (pseudo)
Nambu-Goldstone boson, this $p$-expansion is valid
when $M_{\pi}L \gg 1$. It is well known what happens when
one takes the chiral limit in a volume such that
$M_{\pi}L$ crosses unity and gets even smaller 
\cite{Gasser:1987ah, Neuberger:1987zz}: The propagators of
the pseudo Nambu-Goldstone bosons blow up for one single
momentum mode, the one of zero four-momentum. This invalidates the usual
perturbative expansion, and a different technique is required.
A solution to this problem was given in  \cite{Gasser:1987ah}
in terms of a so-called $\epsilon$-expansion. In this scheme
the zero-momentum mode is, in a sense that becomes more clear below,
integrated out exactly, while all the remaining momentum
modes are treated perturbatively. Since the chiral Lagrangian
involves an infinite series of terms, and since it is only the perturbative
expansion that is jeopardized, ``exact'' integration here refers to 
the term that is leading order in the quark masses $m$. 

The $\epsilon$-expansion is thus perfectly suited for studying 
the extreme case where the quark mass $m$ is so small
that the pion Compton wave length overcomes the 
size of the volume, $M_{\pi}L \ll 1$.
Since the zero-mode becomes dominant in this $\epsilon$-regime, 
physical observables are mostly dependent on the leading
low-energy constants: the infinite-volume chiral condensate $\Sigma$ and
the pion decay constant $F$, both in the chiral limit. The 
next-to-leading order terms (with coefficients $L_i$'s)
at infinite volume are treated in perturbative fashion. This
is similar to the $p$-regime expansion, but the ordering of terms
is different. Also in this respect, the $\epsilon$-regime provides
an intriguing alternative to more conventional ChPT since different
parts of the chiral theory are being probed to any given order.
We note that the studies have now also been extended to
Wilson ChPT where one has more terms which explicitly 
break the symmetry \cite{Shindler:2008ri}.

As the $\epsilon$-regime deals with the extremely chiral limit where 
non-trivial finite-size scaling starts to appear 
(but still far from the symmetric phase 
since one keeps $L\gg 1/\Lambda_{\rm QCD}$), universality is at work.
Perhaps the most important example of this is the equivalence of the
zero-mode or vacuum part of the theory 
to chiral Random Matrix Theory (ChRMT) 
\cite{Shuryak:1992pi, Damgaard:1998xy}. 
Little is known in detail on how these universal phenomena
cease and the ones depending on the dynamics specific to QCD appear 
when the quark mass increases and $M_{\pi}$ becomes
of order $1/L$ \cite{Osborn:1998nf}.
In particular, it is not known precisely
how the spectral density of the Dirac operator, described by ChRMT 
in the low end matches on to the spectrum at larger scales,
in the $p$-regime
\cite{Smilga:1993in}-\cite{Damgaard:1997ye}.

Recent developments in both computational facilities
and algorithms
have allowed simulations of full lattice QCD 
near the chiral limit, but no study has until now reached 
deep inside the $\epsilon$-regime except at rather strong coupling.
Although results have often compared favorably to 
the $\epsilon$-expansion of ChPT,
there may still be large systematic errors due to 
the condition $M_{\pi}L \ll 1$ not being well fulfilled
(see, $e.g.$, ref. \cite{Fukaya:2007fb}).
One might therefore ask whether it is possible to 
have a new approach which smoothly 
connects the $p$-expansion and $\epsilon$-expansion
and which remains valid even in the region $M_{\pi}L\sim 1$.
Recently, steps have been taken in that direction
by means of a so-called mixed expansion
\cite{Bernardoni:2007hi, Bernardoni:2008ei} (see also
ref. \cite{Damgaard:2007ep}), 
where one treats the very light flavors 
with the counting rules of the $\epsilon$-expansion,
while heavier flavors are counted according to the 
$p$-expansion. The results turned out to be
mixtures of the properties of the $\epsilon$
and $p$ regimes: zero-mode fluctuations from the
light sector in addition to 1-loop corrections from
the heavier sector that include 
chiral logs and some of the $L_i$'s. 
But Refs. \cite{Bernardoni:2007hi, Bernardoni:2008ei} treated
the light and heavy flavors separately
and did not attack directly the regime where $M_{\pi}L \sim 1$.
Actually, two regimes at play here: one is the
first obvious threshold when $M_{\pi} \sim 1/L$, the other is
when $M_{\pi} \sim 1/L^2$, the scale of the $\epsilon$-regime.
The question is what happens in-between.

In this paper, we suggest a new perturbative approach
where all the terms in the $p$-expansion are kept but
the zero mode is treated in exactly same way as in
the $\epsilon$-expansion\footnote{We understand that
F. Niedermayer (unpublished) has considered an analogous
scheme in the context of the $O(n)$ sigma model.}. The
expansion thus considers the zero momentum mode on a 
different footing from the rest, partially resumming
terms to all orders. Before reaching the $\epsilon$-regime
this means that an infinite series of terms that are
normally considered in the perturbative expansion are
included to all orders. The result is a slightly re-ordered
perturbation theory expansion that is free from perturbative
infrared singularities in the chiral limit. 
As an example, we compute here a formula for the chiral condensate
which smoothly connects the results of the 
$p$-regime \cite{Bijnens:2006ve} and the $\epsilon$-regime.
We will argue that our formula is reliable even 
in the intermediate region where we go from $M_{\pi} \sim 1/L$
to $M_{\pi} \sim 1/L^2$.

Using the partial quenching technique based on replicas,
we can treat a general theory with valence flavors and
physical sea quark flavors of masses which are non-degenerate.
This allows us to take
the discontinuity on the imaginary axis of the valence quarks,
and thus obtain the corresponding spectral density of
the Dirac operator. It is also given by
a smooth formula that connects known results in the 
$\epsilon$-regime  \cite{Damgaard:1998xy} and $p$-regime 
\cite{Smilga:1993in, Osborn:1998qb}.

The rest of our paper is organized as follows.
In Section \ref{sec:LO}, we describe in detail our new perturbative 
method in ChPT. The chiral condensate to next-leading order
is calculated in Section \ref{sec:condensate}.
Taking the discontinuity on the imaginary axis, we obtain the 
spectral density of the Dirac operator
in Section \ref{sec:s-density}.
We show in Section \ref{sec:connection}
that our results provide a smooth connection between
the $\epsilon$ and $p$ regimes.
In Section \ref{sec:examples} we present some numerical examples  
which are useful when comparing with lattice QCD simulations.
Conclusions and an outlook are given in Section \ref{sec:conclusion}.

\section{A  chiral expansion at finite volume}
\label{sec:LO}

Let us consider $N_f$-flavor chiral perturbation
theory in a finite volume ($V=L^3T$),
\begin{eqnarray}
\mathcal{L}
&=&
\frac{F^2}{4}{\rm Tr}
[\partial_\mu U(x)^\dagger\partial_\mu U(x) ]
-\frac{\Sigma}{2}{\rm Tr}
[\mathcal{M}^{\dagger}e^{-i\theta /N_f}U(x)
+U(x)^\dagger e^{i\theta /N_f}\mathcal{M}]+\cdots,
\end{eqnarray}
where $U(x)\in SU(N_f)$ and $\theta$ denotes 
the vacuum angle.
Here, $\Sigma$ is the chiral condensate
and $F$ denotes the pion decay constant 
both in the chiral limit.
There are of course next-to-leading order terms,
indicated here by ellipses, with additional low-energy
constants denoted by $L_i$'s, $H_i$'s and beyond.

In the partially quenched case, we use the replica method
where the calculations are done within an 
($N_f+N_v+(N-N_v)$)-flavor theory followed by 
the replica limit $N \to 0$ \cite{Damgaard:2000gh, Damgaard:2001js}.
The ordinary physical $N_f$-theory result can clearly be 
viewed either as one where
$N = N_v = 0$, or, alternatively, one where 
$m_v=m_f$ with $m_f$ denoting one of the physical quark masses.

From now on we consider sectors of fixed topology $\nu$, 
obtained by Fourier transforming in $\theta$ in the usual way. This
extends our integration from $SU(N+N_f)$ to $U(N+N_f)$
in the zero-momentum sector.

For the mass matrix, we consider a general diagonal case,
\begin{eqnarray}
\mathcal{M} &=& 
\mathrm{diag}(\underbrace{m_{v},m_{v},\cdots}_{N},
\underbrace{m_1,m_2,\cdots}_{N_f}),
\end{eqnarray}
where we have $N = N_v + (N-N_v)$ replicated flavors 
and $N_f$ physical flavors. 
Here $N_v$ is the number of,
in this case, degenerate valence quarks. What we do below
can straightforwardly be generalized to non-degenerate
valence quarks by just adding copies of each.

We start by factorizing the fields into 
the zero-momentum mode $U$ and non-zero modes $\xi(x)$,
\begin{eqnarray}
\label{eq:param}
U(x)=U\exp(i\sqrt{2}\xi(x)/F),
\end{eqnarray}
and expand perturbatively in $\xi(x)$ just as in the 
$\epsilon$-regime \cite{Gasser:1987ah}.
But here we give the same counting rules
for the fields and other parameters as in the $p$-regime:
\begin{eqnarray}
\partial_\mu \sim {\cal O}(p),\;\;\;
\xi(x) \sim {\cal O}(p),\;\;\;
\mathcal{M}\sim {\cal O}(p^2),\;\;\;
T, L \sim {\cal O}(1/p), \label{pcounting}
\end{eqnarray}
in units of the cut-off $4\pi F$.
The aim is to see if we can tune quark masses so
that we go from mass scales ${\mathcal M} \sim {\cal O}(1/L^2)$
through ${\mathcal M} \sim {\cal O}(1/L^4)$ to zero.
Here we of course assume that the linear sizes of the volume, $L$ and $T$, 
are much larger than  the inverse QCD scale $\Lambda_{QCD}$
so that the effective theory is valid.

The above parametrization Eq.~(\ref{eq:param}) leads to a well-known
Jacobian in the functional integral measure \cite{Hansen:1990un}.
Although it is easily taken into account,
its contribution is ${\cal O}(p^6)$ and beyond the
accuracy with which we do actual calculations in this paper.

We now expand the Lagrangian in $\xi(x)$ according to the
$p$-counting Eq.~(\ref{pcounting}), 
and write down the terms relevant to one-loop order
for the chiral condensate,
\begin{eqnarray}
\label{eq:Lchpt}
{\cal L} &=& -\frac{\Sigma}{2}\mathrm{Tr}\left[
\mathcal{M}^\dagger U+U^\dagger \mathcal{M}\right]
+\frac{1}{2}\mathrm{Tr}(\partial_\mu \xi)^2
+\frac{1}{2}\sum_i M^2_{ii}[\xi^2]_{ii}
\nonumber\\&&
+\frac{\Sigma}{2F^2}
{\rm Tr}[\mathcal{M}^\dagger (U-1)\xi^2+\xi^2(U^\dagger-1)\mathcal{M}]
\nonumber\\&&
-L_6 \left(\frac{2\Sigma}{F^2}
{\rm Tr}
[\mathcal{M}^{\dagger}U
+U^\dagger \mathcal{M}]\right)^2
\nonumber\\
&&-L_7 \left(\frac{2\Sigma}{F^2}
{\rm Tr}
[\mathcal{M}^{\dagger}U
-U^\dagger \mathcal{M}]\right)^2
\nonumber\\
&&-\frac{L_8}{2} \left(\frac{2\Sigma}{F^2}\right)^2
{\rm Tr}
[
(\mathcal{M}^{\dagger}U
+U^\dagger \mathcal{M})^2
+(\mathcal{M}^{\dagger}U
-U^\dagger \mathcal{M})^2
]
\nonumber\\
&&-H_2\left(\frac{2\Sigma}{F^2}\right)^2
{\rm Tr}
[\mathcal{M}^{\dagger}\mathcal{M}]
+\cdots, \label{LNLO}
\end{eqnarray}
where $M^2_{ij}=(m_i+m)\Sigma/F$ and $L_i$'s are the usual
higher-order low-energy constants of ChPT. Here we have added and
subtracted a conventional mass term of the $p$-regime.
We will treat the mass term of
the first line in eq. (\ref{LNLO}) as part of the exact
Gaussian integration that leads to the conventional massive propagator
of the $p$-regime, while the remaining terms are treated in
a perturbative expansion. We return to this point below.
The contact term $H_2$ has no direct physical significance,
but it is needed as a counterterm for the one-loop correction
to the condensate \cite{Gasser:1987ah}.
All linear terms in $\xi$ are absent due to
\begin{eqnarray}
\int d^4 x\; \xi(x) =0 .
\end{eqnarray}

In Eq.~(\ref{LNLO}) the first line contains terms that in the
usual $p$-expansion are of order $p^2$ (the first,
a trivial constant in the usual infinite-volume $p$-expansion) and $p^4$
(the remaining two). In the $\epsilon$-expansion the first two terms
on the same line are of order $\epsilon^4$, while the third
is of order $\epsilon^6$. In the last four lines we have written
out explicitly those terms that are of order $p^4$ in the usual 
infinite-volume $p$-expansion (but trivial constants there). In the 
$\epsilon$-counting these terms are of order $\epsilon^8$.
Other terms involving
the $L_i$'s will be of order $p^6$ in that same counting. However,
as with the measure term, these terms will not contribute to the
chiral condensate that we will compute below.

The one single term we have not yet discussed is that of
\begin{eqnarray}
\Delta {\cal L} ~=~ \frac{\Sigma}{2F^2}
{\rm Tr}[\mathcal{M}^\dagger (U-1)\xi^2+\xi^2(U^\dagger-1)\mathcal{M}] ~.
\label{U-1}
\end{eqnarray}
In the usual $p$-expansion a term of this type first occurs at
order $p^5$ (because there will be three powers of $\xi$), and in 
the $\epsilon$-expansion it is of order $\epsilon^6$.
Here we treat any matrix elements of 
$\mathcal{M}^\dagger (U-1)$ (and its complex conjugate
$(U^\dagger-1)\mathcal{M}$) as of  
${\cal O}(p^3)$ for all values of $\mathcal{M}$.
By performing the exact group integration over $U$, 
we can check that the combination $\mathcal{M}^\dagger (U-1)$ 
gives NLO contributions ($\lesssim {\cal O}(p^3)$) 
to the results. We illustrate this in Section 
 \ref{sec:connection}. 
In Appendix \ref{app:zero-mode-first}
we describe an alternative method which expresses the magnitude
of the contribution from the term in eq. ({\ref{U-1}) directly in
terms of masses and the volume $V$, thus giving a precise counting
of this term in terms of $p \sim 1/L$. This alternative method gives
identical results, but is in practice more cumbersome than the
scheme presented here.

By taking $\mathcal{M}^\dagger (U-1)\sim {\cal O}(p^3)$,
one can thus, to this order, rewrite the Lagrangian
\begin{eqnarray}
\label{eq:Lchpt2}
{\cal L} &=& -\frac{\Sigma}{2}\mathrm{Tr}\left[
\mathcal{M}^\dagger U+U^\dagger \mathcal{M}\right]
+\frac{1}{2}\mathrm{Tr}(\partial_\mu \xi)^2
+\frac{1}{2}\sum_i M^2_{ii}[\xi^2]_{ii}
\nonumber\\&&
+\frac{\Sigma}{2F^2}
{\rm Tr}[\mathcal{M}^\dagger (U-1)\xi^2+\xi^2(U^\dagger-1)\mathcal{M}]
\nonumber\\&&
-L_6 \left(\frac{2\Sigma}{F^2}
{\rm Tr}
[\mathcal{M}^{\dagger}U
+U^\dagger \mathcal{M}]\right)^2
-\frac{2L_8+H_2}{4} \left(\frac{2\Sigma}{F^2}\right)^2
{\rm Tr}
[
(\mathcal{M}^{\dagger}U
+U^\dagger \mathcal{M})^2
],
\end{eqnarray}
where the second line is treated as an NLO interaction term
and the contribution from the $L_7$-term has been dropped.

It should be stressed at this point that adding and subtracting
an ordinary $p$-regime mass term and then expanding the term
(\ref{U-1}) perturbatively has the effect or a complete
re-ordering and partial resummation of the perturbative
series. This resummation comes from the fact that when the
chiral limit is taken and $M_{\pi}^2$ is no longer of
order $p^2$, but smaller, we still keep the full massive
propagator. The error in doing this, rather than expanding
the propagator to the needed order in $M_{\pi}^2$, is however
always of yet higher order and part of the unavoidable uncertainty in
any fixed-order perturbative calculation. We always keep the
full massive propagator in the expressions and plots we present below.

The Feynman rule for the $\xi$-propagator is thus 
obtained as usual, except that
the zero-momentum modes are not included:
\begin{eqnarray}
\langle \xi_{ij}(x)\xi_{kl}(y)\rangle_\xi
&=&\delta_{il}\delta_{jk}\bar{\Delta}(x-y, M^2_{ij})
-\delta_{ij}\delta_{kl}\bar{G}(x-y,M^2_{ii},M^2_{kk}),
\end{eqnarray}
and the second term comes from the
constraint ${\rm Tr} \xi=0$.
The propagators $\bar{\Delta}$ and $\bar{G}$ are given 
by\footnote{We do not consider the fully quenched theory in this
paper. We thus have $N_f \neq 0$ in all that follows.},
\begin{eqnarray}
\bar{\Delta}(x,M^2)&=&\frac{1}{V}\sum_{p\neq 0}
\frac{e^{ipx}}{p^2+M^2},\\
\bar{G}(x,M_{ii}^2,M^2_{jj})&=&
\frac{1}{V}\sum_{p\neq 0}\frac{e^{ipx}}{(p^2+M_{ii}^2)(p^2+M_{jj}^2)
\left(\sum^{N_f}_f\frac{1}{p^2+M^2_{ff}}
\right)},
\end{eqnarray}
where the summation is taken over the non-zero 4-momentum
\begin{eqnarray}
p=2\pi(n_t/T, n_x/L, n_y/L, n_z/L),
\end{eqnarray}
with integer $n_\mu$.


\section{The chiral condensate}
\label{sec:condensate}



The chiral condensate of a valence flavor is obtained 
in the conventional manner 
by adding a source to the mass matrix; $\mathcal{M}\to\mathcal{M+J}$
and differentiating the partition function 
$\mathcal{Z}(\mathcal{M+J})$ with respect to $\mathcal{J}$.
To leading order in our expansion this gives
\begin{eqnarray}
\label{eq:1-loop-calculation}
\langle \bar{q}_vq_v \rangle_{\nu}^{LO}
&\equiv& \left.\frac{1}{V}\frac{\partial}{\partial \mathcal{J}_{vv}} 
\ln \mathcal{Z} (\mathcal{M+J})\right|_{\mathcal{J}=0}
=
\frac{\Sigma}{2}
\left\langle U_{vv}+U^\dagger_{vv}\right\rangle_U,
\end{eqnarray}
where the zero-mode integral 
$\left\langle U_{vv}+U^\dagger_{vv}\right\rangle_U$
is computed non-perturbatively with respect to the zero-mode partition
function
\begin{eqnarray}
{\cal{Z}}^{\nu}_{LO} ~=~
\int_{U(N+N_f)}\!dU~ (\det U)^{\nu}
\exp\left[\frac{\Sigma}{2}\mathrm{Tr}\left(
\mathcal{M}^\dagger U+U^\dagger \mathcal{M}\right)\right].
\label{Znuzero}
\end{eqnarray}

The analytical formula is known
for the most general partially quenched case 
with non-degenerate physical $N_f$-flavors \cite{Splittorff:2002eb}.
Some details are summarized in Appendix \ref{app:zero-mode}. Here
we simply {\em define}
\begin{eqnarray}
\label{eq:cond}
\hat{\Sigma}_\nu^{\rm PQ}(\mu_v, \{\mu_{sea}\}) ~\equiv~
\frac{1}{2}\left\langle U_{vv}+U^\dagger_{vv}\right\rangle_U ~,
\end{eqnarray}
where $\mu_v = m_v \Sigma V$ and the set of
the dynamical flavors are denoted by 
$ \{\mu_{sea}\}=\{\mu_1,\mu_2,\cdots \}$ with $\mu_i=m_i \Sigma V$.

At next-to-leading order, it is convenient to first 
calculate the 1-loop perturbative correction 
due to the non-zero modes.
This can be done by simply evaluating
\begin{eqnarray}
\left\langle 1-\int d^4 x \frac{\Sigma}{2F^2}
{\rm Tr}[(\mathcal{M+J})^\dagger (U-1)\xi^2+\xi^2(U^\dagger-1)(\mathcal{M+J})]
\right\rangle_\xi, 
\end{eqnarray}
where $\langle\cdots \rangle_\xi$ denotes the integral over $\xi$,
and then re-exponentiating it. The effective 
Lagrangian (with a scalar source $\mathcal{J}$) then reads
\begin{eqnarray}
\label{eq:Lchpt3}
{\cal L}_{\rm eff}(\mathcal{J}) &=& 
-\frac{\Sigma}{2}\sum_i Z_i
\left[
(\mathcal{M+J})^\dagger U+U^\dagger (\mathcal{M+J})\right]_{ii}
\nonumber\\&&
+\frac{1}{2}\mathrm{Tr}(\partial_\mu \xi)^2
+\frac{\Sigma}{F^2}{\rm Tr}[(\mathcal{M+J})(\xi^2-\langle\xi^2 \rangle_\xi)],
\end{eqnarray}
where
\begin{eqnarray}
\label{eq:Z_i}
Z_i \equiv 1-\frac{1}{F^2}\left[
\sum_j \bar{\Delta}(0,M^2_{ij})-\bar{G}(0,M^2_{ii},M^2_{ii})
-16L_6\sum_j M^2_{jj}-4(2L_8+H_2)M^2_{ii}
\right].
\end{eqnarray}

Both of the 1-loop integrations $\bar{\Delta}(0,M^2)$ and
$\bar{G}(0,M^2,M^2)$ are UV divergent and their divergences are
absorbed into the bare parameters $L_6$, $L_8$ and $H_2$.
With an appropriate regularization such as dimensional regularization
(see ref. \cite{Hasenfratz:1989pk} for a discussion of this issue)
$\bar{\Delta}(0,M^2)$ is given by
\begin{eqnarray}
\bar{\Delta}(0,M^2) = \frac{M^2}{16\pi^2}
(\ln M^2 + c_1)+ \bar{g_1}(M^2), 
\end{eqnarray}
where $c_1$ represents the conventional logarithmic divergence
which is independent of $M$ and the volume \cite{Hasenfratz:1989pk}.
The function $\bar{g}_1$ represents the finite size effects,
\begin{eqnarray}
\label{eq:g1}
\bar{g}_1(M^2)\equiv \frac{1}{(4\pi)^2}
\int_0^\infty \sum_{a \neq 0}\frac{dr}{r^2}
\exp\left(-r M^2-
\frac{1}{4r}\sum_\mu (a_\mu)^2\right)-\frac{1}{M^2V},
\end{eqnarray}
and the sum is taken over a 4-dimensional vector $a_\mu=n_\mu L_\mu$
($L_\mu$ being the lattice size in the $\mu$-th direction)
with integer $n_\mu$.
It is particularly important to note that the massless limit 
$M^2\to 0$ of $\bar{g}_1$ is finite and given by
$\bar{g}_1(0)=-\beta_1/\sqrt{V}$ where $\beta_1$ is 
the so-called shape coefficient \cite{Hasenfratz:1989pk}.
This term dominates
the NLO correction in the $\epsilon$-regime.
A detailed numerical treatment of $\bar{\Delta}(0,M^2)$ and 
its derivative is discussed in Appendix \ref{app:g1g2}.

Expanding $\bar{G}(0,M^2,M^2)$ in $M^2$ 
(see Appendix \ref{app:G-property}), we get
\begin{eqnarray}
\bar{G}(0,M^2,M^2) = \frac{2}{N_f}\bar{\Delta}(0,M^2)
-\frac{1}{N_f^2}\sum_j \bar{\Delta}(0,M_{jj}^2) + \cdots,
\end{eqnarray}
where the sum $\sum_j$ is taken over physical flavors only.
Since the ${\cal O}(M^4)$ terms indicated by ellipses are
UV finite, the divergence in $Z_i$, 
Eq.~(\ref{eq:Z_i}), is in total, 
\begin{eqnarray}
Z_i\sim -\frac{1}{16\pi^2F^2}
\left[\sum_j \frac{M^2_{ii}+M^2_{jj}}{2} - 
\frac{2}{N_f}M^2_{ii} + \frac{1}{N_f^2}\sum_j M^2_{jj}\right]
c_1,
\end{eqnarray}
which can be absorbed into redefinitions of 
$L_6$, $L_8$ and $H_2$,
\begin{eqnarray}
L_6 &=& \frac{1}{(16\pi)^2}
\left(\frac{1}{2}+\frac{1}{N^2_f}\right)
c_1
+L_6^r,\;\;\;
L_8 = \frac{1}{(16\pi)^2}
\left(\frac{N_f}{2}-\frac{2}{N_f}\right)
c_1
+L_8^r,\nonumber\\
H_2 &=& \frac{2}{(16\pi)^2}
\left(\frac{N_f}{2}-\frac{2}{N_f}\right)
c_1
+H_2^r,
\end{eqnarray}
where the renormalized constants are denoted by 
$L_6^r$, $L_8^r$ and $H^r_2$.
This renormalization is identical to that 
of the infinite volume case \cite{Gasser:1987ah},
as it should be.
\\

To this order,
the chiral condensate at fixed topology can thus
be written
\begin{eqnarray}
\label{eq:condNLO}
\langle \bar{q}_vq_v \rangle_{\nu}
&=&
\Sigma \frac{\mu^\prime_v}{\mu_v}\hat{\Sigma}^{\rm PQ}_\nu
(\mu^\prime_v, \{\mu^\prime_{sea}\}) ~,
\end{eqnarray}
where $\mu^\prime_v=Z_v m_v \Sigma V$ and
$\{\mu^\prime_{sea}\}=\{Z_1 m_1 \Sigma V, Z_2 m_2 \Sigma V, \cdots\}$. 
Note that the arguments of the function now include
the chiral logarithms as explicitly seen in $Z_i$ (see Eq.~(\ref{eq:Z_i})).


The expression in Eq.~(\ref{eq:condNLO}) 
looks simple and compact. But in order to see the valence mass 
dependence, it is more convenient to decompose $Z_v$ 
into two finite pieces,
\begin{eqnarray}
Z_v &=& Z^0_v +\delta Z_v(m_v),\\
Z^0_v &\equiv & 1-\frac{1}{F^2}\left[
\sum_j \bar{\Delta}(0,M^2_{jj}/2)-\bar{G}(0,0,0)
-16L_6\sum_j M^2_{jj}\right],\\
\delta Z_v(m_v)&\equiv &
-\frac{1}{F^2}\left[
\sum_j (\bar{\Delta}(0,M^2_{jv})
-\bar{\Delta}(0,M^2_{jj}/2))
\right.\nonumber\\
&&\left.\hspace{0.5in}
-(\bar{G}(0,M^2_{vv},M^2_{vv})-\bar{G}(0,0,0))
-4(2L_8+H_2)M^2_{vv}\right],
\end{eqnarray}
where only $\delta Z_v(m_v)$ has a valence mass dependence,
and it vanishes in the limit $m_v\to 0$.
Note that $\bar{G}(0,0,0)$ is infra-red finite 
(see Appendix \ref{app:G-property}).

With the above decomposition, the condensate can be
expressed as
\begin{eqnarray}
\label{eq:condNLO-expand}
\langle \bar{q}_vq_v \rangle_{\nu}
&=&
\Sigma \left[Z^0_v\;
\hat{\Sigma}^{\rm PQ}_\nu(Z^0_v\mu_v, \{\mu^\prime_{sea}\})
\right.\nonumber\\&&\left.
+\delta Z_v(m_v)\left(
\hat{\Sigma}^{\rm PQ}_\nu(\mu_v, \{\mu_{sea}\})
+\mu_v\frac{\partial}{\partial \mu_v}
\hat{\Sigma}^{\rm PQ}_\nu(\mu_v, \{\mu_{sea}\})
\right)+\cdots\right]
\nonumber\\
&=&
\Sigma \left[Z^0_v\;
\hat{\Sigma}^{\rm PQ}_\nu(Z^0_v\mu_v, \{\mu^\prime_{sea}\})
+\delta Z_v(m_v)\right]+{\cal O}(p^4) ~,
\end{eqnarray}
where $\mu_v = m_v \Sigma V$.
In the second line we have used the fact 
that $\delta Z_v(m_v)\sim {\cal O}(M^2_{vv})$
does not contribute until 
$\hat{\Sigma}^{\rm PQ}_\nu(\mu_v, \{\mu^\prime_{sea}\})$
becomes close to $1+{\cal O}(p^2)$.
By the same technique, one can replace the
sea quark's argument $\mu_i^\prime = Z_i m_i \Sigma V$
by  $Z_i^0 m_i \Sigma V$ without producing no additional term.
The explicit form will be given in Section 6.


It is now clear that the chiral condensate near the
chiral limit is dominated by zero modes, and hence expressed through
combinations of Bessel functions as in the $\epsilon$-regime. 
The argument
$Z^0_v m_v \Sigma V$, however, includes
the chiral logarithm of the sea quarks in $Z^0_v$.
As the valence mass increases, 
$\hat{\Sigma}^{\rm PQ}_\nu(\mu_v, \{\mu_{sea}\})$ approaches
unity and the ordinary valence quark chiral logarithm appears in 
$\delta Z_v(m_v)$.
For yet larger values of $m_v$\footnote{Because $m_v = \mu_v/(\Sigma L^4)$,
such a term first appears at NNLO in the pure 
$\epsilon$-expansion.}, the term proportional to
$ 2L^r_8 +H^r_2$ 
becomes important. The unphysical quantity $H^r_2$ depends 
on the regularization scheme, and the condensate is
then not unambiguously defined, as is well known \cite{Gasser:1983yg}.
In such a region, one has to eliminate the $H^r_2$ dependence 
to obtain unambiguous physical observables. An example would be to 
consider a difference between two topological sectors.
As we will see in the next section, the spectral density is 
also free from this ambiguity.

\section{The spectral density of the Dirac operator}
\label{sec:s-density}

In the previous discussion, we assumed that 
all quark masses were real and positive. 
As is well-known \cite{Damgaard:1998xy}, by considering 
the expressions for imaginary valence quark masses,
one can calculate all spectral correlation functions
and individual eigenvalues distributions 
of the Dirac operator of the $N_f$-flavor theory.
In this case partial quenching is simply used to extract
a physical observable in the full theory.
One expresses the valence quark condensate as a spectral
sum over the
Dirac eigenvalues $i\lambda_k$'s ($\lambda_k$ is real),
\begin{eqnarray}
\langle \bar{q}_vq_v
\rangle|_{m_v}=\frac{1}{V}\sum_k 
\left\langle \frac{1}{m_v+i\lambda_k}\right\rangle.
\label{cond-rho}
\end{eqnarray}
Since every non-zero eigenvalue comes paired with
one of opposite sign, the condensate satisfies
\begin{eqnarray}
\langle \bar{q}_vq_v\rangle|_{m_v^*} 
= (\langle \bar{q}_vq_v \rangle|_{m_v})^*,\;\;\;
\langle \bar{q}_vq_v\rangle|_{-m_v} 
= -\langle \bar{q}_vq_v \rangle|_{m_v},
\end{eqnarray}
where $*$ denotes complex conjugation.

Using the above, the spectral density at fixed topology $\nu$
is given by
\begin{eqnarray}
\label{eq:general-rho_nu}
\rho_\nu(\lambda) &~\equiv~& \frac{1}{V}
\sum_k \left\langle \delta(\lambda+\lambda_k)\right\rangle_\nu
\nonumber\\
&=& -\frac{1}{2\pi V} \sum_k 
\lim_{\epsilon\to 0}\left\langle \frac{1}{i(\lambda+\lambda_k)-\epsilon}
- \frac{1}{i(\lambda+\lambda_k)+\epsilon}\right\rangle_\nu
\nonumber\\
&=&\lim_{\epsilon \to 0}
\frac{1}{2\pi}(\langle \bar{q}_vq_v
\rangle_\nu|_{m_v=i\lambda-\epsilon}-
\langle \bar{q}_vq_v\rangle_\nu|_{m_v=i\lambda+\epsilon})
\nonumber\\
&=& \frac{\Sigma}{\pi} \left[
Z^0_v\;{\rm Re}\hat{\Sigma}^{\rm PQ}_\nu
(i\lambda \Sigma V Z^0_v, \{\mu^\prime_{sea}\})
+{\rm Re}\left(\delta Z_v(i\lambda)\right)\right],\label{rhodisc}
\end{eqnarray}
where we have neglected the $\delta(\lambda)$ term 
which represents the exactly zero eigenvalues 
due to the non-trivial topological charge $\nu$.
Similar expressions exist for all higher spectral correlation
functions.

For the calculation of ${\rm Re}\left(\delta Z_v(i\lambda)\right)$,
we need the discontinuities of various functions. 
For the log-terms, for example, one obtains
\begin{eqnarray}
\label{eq:analcont2}
{\rm Re}  \ln (m+\sqrt{m^2_v})|_{m_v=i\lambda} 
&=& \frac{1}{2}  \ln (m^2+\lambda^2),\\
{\rm Im}  \ln (m+\sqrt{m^2_v})|_{m_v=i\lambda} 
&=&  \arctan \frac{\lambda}{m},
\end{eqnarray}
where $m$ denotes a real and positive mass, 
for which the limit $m\to 0$ is well-defined.
Some of the other functions occurring in eq. (\ref{rhodisc}) need
to be treated numerically, such as those in $\bar{g}_1$. We collect
some representations useful for numerical purposes in Appendix 
\ref{app:g1g2}.
Note that the term
proportional to $2L_8+H_2$ has disappeared upon
taking the discontinuity across the imaginary valence quark axis\footnote{
This result may seem to contradict the
fact that the condensate, conversely, should follow from a spectral
sum involving $\rho(\lambda)$ as in Eq.~(\ref{cond-rho}). The 
problem is that the spectral sum is UV divergent. It is this divergence
that gives rise to an ambiguity such as indicated by the $2L_8+H_2$-term.}.

Next, using the properties of Bessel functions,
one sees that the first term of Eq.~(\ref{eq:general-rho_nu})
reproduces the known form of the leading contribution to the
microscopic spectral density in the $\epsilon$-regime,
\begin{eqnarray}
{\rm Re}\hat{\Sigma}^{\rm PQ}_\nu(i\zeta, \{\mu_{sea}\})
= \pi \hat{\rho}^{mic}_\nu (\zeta, \{\mu_{sea}\}),
\end{eqnarray}
where $\hat{\rho}^{mic}_\nu (\zeta, \{\mu_{sea}\})$ is given by
\cite{Damgaard:1997ye}
\begin{eqnarray}
\hat{\rho}^{mic}_\nu (\zeta, \{\mu_{sea}\})
\equiv 
\left|\frac{|\zeta|}{2\prod^{N_f}_f(\zeta^2 + \mu^2_f)}
\frac{\det \tilde{\mathcal{B}}}{\det \mathcal{A}}\right| .
\end{eqnarray}
Here the $N_f\times N_f$ matrix $\mathcal{A}$ and the
$(N_f+2)\times (N_f+2)$ matrix $\tilde{\mathcal{B}}$ are defined by
\begin{eqnarray}
\mathcal{A}_{ij}&=& \mu_i^{j-1}I_{\nu+j-1}(\mu_i)\\
\tilde{\mathcal{B}}_{1j} &=&  \zeta^{j-2}J_{\nu+j-2}(\zeta),\;\;\;
\tilde{\mathcal{B}}_{2j} =  \zeta^{j-1}J_{\nu+j-1}(\zeta),\nonumber\\
\tilde{\mathcal{B}}_{ij} &=&  (-\mu_{i-2})^{j-1}I_{\nu+j-1}(\mu_{i-2})
\;\;\;(i\neq 1,2).
\end{eqnarray}

The general formula for the spectral density can thus conveniently be 
written in a representation suitable for small eigenvalues that go into
the ``bulk'' region, 
\begin{eqnarray}
\label{eq:general-rho_nu2}
\rho_\nu(\lambda) 
&=& \frac{\Sigma}{\pi} \left[
\pi Z^0_v\;\hat{\rho}^{mic}_\nu (Z^0_v\lambda \Sigma V, \{\mu^\prime_{sea}\})
+{\rm Re}\left(\delta Z_v(i\lambda)\right)\right] ~.
\end{eqnarray}

\section{From the $p$-regime to the $\epsilon$-regime}
\label{sec:connection}

In this section we explain how our formulae are 
consistent with known results 
in both $\epsilon$ and $p$ regimes.
We next consider the validity of our expressions in the
intermediate region.

\subsection{Checks in the $p$-regime and limit to the $\epsilon$-regime}

We first check that the expression derived in the previous
section reproduces
known results in the conventional perturbative 
$p$-expansion.
The exact zero-mode integral above is expressed by
complicated combinations of 
Bessel functions. But for large $m_i \Sigma V$,
these Bessel functions can be expanded in $1/m_i \Sigma V$ 
(see, $e.g.$, ref. \cite{Damgaard:2000di} for details),
\begin{eqnarray}
\label{eq:plimitSigma}
\hat{\Sigma}^{\rm PQ}_\nu(\mu_v,\{\mu_{sea}\})
&=&
1-\sum_j \frac{1}{\mu_v+\mu_j}
+\frac{4\nu^2 -1}{8\mu_v^2}+{\cal O}(p^6),
\end{eqnarray}
which, after summing over topology, gives
\begin{eqnarray}
\label{eq:sigma-pert}
\hat{\Sigma}^{\rm PQ}(\mu_v,\{\mu_{sea}\})
&=& 1-\sum_j \frac{1}{\mu_v+\mu_j}
+\frac{\langle \nu^2\rangle}{2\mu_v^2}+{\cal O}(p^4)
\nonumber\\
&=& 1-\sum_j \frac{1}{\mu_v+\mu_j}
+\frac{1}{2\mu_v^2(\sum_f 1/\mu_f)}+{\cal O}(p^4),
\end{eqnarray}
where we have used
\begin{eqnarray}
\langle \nu^2\rangle = \frac{1}{\sum_f 1/\mu_f} + \cdots.
\end{eqnarray}
in the chiral limit.
Substituting Eq.~(\ref{eq:sigma-pert}) into Eq.~(\ref{eq:condNLO}),
we reproduce the perturbative $p$-regime result,
\begin{eqnarray}
\label{eq:condNLO-pert}
\langle \bar{q}_vq_v \rangle
&=&
\Sigma \left[1-\frac{1}{F^2}\left(
\sum_j \Delta(0,M^2_{vj})-G(0,M^2_{vv},M^2_{vv})
\right)
\right.\nonumber\\ && \hspace{1in}\left.
+\frac{1}{F^2}(16L_6\sum_i M^2_{ii}+4(2L_8+H_2)M^2_{vv})\right].
\end{eqnarray}
As is known from the matching between $\epsilon$ and $p$ regimes,
the zero-mode fluctuations which give rise to the 
second and third terms in Eq.~(\ref{eq:sigma-pert}) 
are absorbed in the momentum sum in $\bar{\Delta}$ and
$\bar{G}$ so that we recover the usual propagators
\begin{eqnarray}
\Delta(x,M^2)&=&\frac{1}{V}\sum_{p}
\frac{e^{ipx}}{p^2+M^2},\\
G(x,M_{ii}^2,M^2_{jj})&=&
\frac{1}{V}\sum_{p}\frac{e^{ipx}}{(p^2+M_{ii}^2)(p^2+M_{jj}^2)
\left(\sum^{N_f}_f\frac{1}{p^2+M^2_{ff}}
\right)},
\end{eqnarray}
of the ordinary $p$-expansion.

The above expression agrees 
with known results that can be found in, $e.g.$, 
the work of Osborn {\it et al.} \cite{Osborn:1998qb}.
They derived a formula
for the partially quenched case with $N_f$ 
degenerate flavors of mass $M^2_{sea}$ in which case one can use
\begin{eqnarray}
G(x,M_{vv}^2,M^2_{vv})=
\frac{1}{N_f}\left[\Delta(x,M^2_{vv})
+(M^2_{vv}-M^2_{sea})\partial_{M^2_{vv}}\Delta(x,M^2_{vv})\right].
\end{eqnarray}

One can also check the $\epsilon$-regime results 
at fixed topology $\nu$ are precisely reproduced 
just by reducing the quark masses in the formula
Eq.~(\ref{eq:condNLO-expand}). One  notes that in that limit,
\begin{eqnarray}
Z_v^0 &\to& 1+\frac{1}{F^2}\left(\frac{N_f^2-1}{N_f}\right)
\frac{\beta_1}{\sqrt{V}},\;\;\;\;\;\;
\delta Z_v(m_v) \to 0.
\end{eqnarray}\\

The spectral density of the Dirac operator
can easily be compared to known results in different limits
as well.
First of all, the Banks-Casher relation is trivially reproduced
when we take the limit $V\to \infty$ before $m_i \to 0$,
\begin{eqnarray}
\label{eq:Banks-Casher}
\lim_{m_i \to 0}\lim_{V\to \infty}
\rho_{\nu}(\lambda=0) &=& \frac{\Sigma}{\pi},
\end{eqnarray}
after having used 
$\hat{\rho}^{mic}_{\nu} (\infty, \{\mu_{sea}\})=1/\pi$,
$Z^0_v|_{V\to \infty, m_i\to 0} \to 1$, and $\delta Z_v(0)=0$.
In the same limit above, but for finite $\lambda$, one obtains
\begin{eqnarray}
\label{eq:Smilga-Stern}
\lim_{m_i \to 0}\lim_{V\to \infty}
\rho_{\nu}(\lambda) &=& \frac{\Sigma}{\pi}
\left[1+{\rm Re}\delta Z_v(i\lambda)|_{V \to \infty, m_i\to 0}\right]\nonumber\\
&=&\frac{\Sigma}{\pi}\left[1+\left(\frac{N_f^2-4}{N_f}\right)
\frac{\Sigma}{32\pi F^4}|\lambda|\right],
\end{eqnarray}
which is the result of Smilga and Stern \cite{Smilga:1993in}.
If one keeps the sea quark masses $m_i=m$ finite (and degenerate),
one gets
\begin{eqnarray}
\lim_{V\to \infty}
\rho_{\nu}(\lambda) &=& \frac{\Sigma}{\pi}
\left[Z^v_0|_{V \to \infty}+
{\rm Re}\delta Z_v(i\lambda)|_{V \to \infty}\right]\nonumber\\
&=&\frac{\Sigma}{\pi}\left[1+
\frac{\Sigma}{32\pi^2N_f F^4}\left(
2N_f^2 |\lambda|\arctan \frac{|\lambda|}{m}-4\pi|\lambda|
\right.\right.\nonumber\\
&&\left.\left.-N_f^2 m \ln \frac{\Sigma^2}{F^4}\frac{m^2+\lambda^2}{\mu_{sub}^4}
-4m \ln \frac{\Sigma}{F^2}\frac{|\lambda|}{\mu_{sub}^2}\right)+
\frac{32 N_f L^r_6 (\mu_{sub})\Sigma m}{F^4}\right],
\end{eqnarray}
where $\mu_{sub}$ denotes the subtraction scale. This is consistent
with the formula by Osborn {\it et al.} \cite{Osborn:1998qb}.
In the case of finite $V$ and 
very small $m_i$ and $\lambda$, the general result in the $\epsilon$-regime 
\cite{Damgaard:1997ye},
\begin{eqnarray}
\rho_{\nu}(\lambda) &=& \Sigma 
\hat{\rho}^{mic}_\nu(\lambda\Sigma V, {\mu_{sea}}),
\end{eqnarray}
is easily recovered upon noting that to leading order we have $Z_v^0 = 1$.

\subsection{Intermediate regime}

As seen in the above discussion, 
our formulae for the condensate and the spectral density
smoothly connect the results in the $\epsilon$-regime
with those of the $p$-regime. But we also need to know
the precision in the intermediate region.
The $\epsilon$-regime assumes $M_{vv} L \ll 1$
while the $p$-regime counting requires $M_{vv}L \gg 1$.
Our re-ordered perturbative expansion has removed this constraint.
Instead, we make use of a non-trivial prescription for
the fourth term in Eq.~(\ref{eq:Lchpt})
where we take $\mathcal{M}^\dagger(U-1)$ to be always
small, specifically of ${\cal O}(p^3)$ or smaller.

One should note that non-zero mode's contributions are 
free from infra-red divergences by construction. This
is also seen explicitly in the finite chiral limit of 
of the finite-volume function $\bar{g}(M^2)$.
For the non-zero modes, there is no need to distinguish 
the $\epsilon$-regime from the $p$-regime.
The smaller the quark masses, the better convergence of 
the non-zero mode expansion.

Therefore, the accuracy of our calculation needs to be assessed
by considering the zero-mode integrals.
To show the general validity of the method,
we need to confirm by explicit evaluation of the group integrals
that the operator $\mathcal{M}^\dagger(U-1)$ consistently
can be taken to be of ${\cal O}(p^3)$ 
or smaller. This in any combination 
of matrix elements, power and for an arbitrary
choice of the mass matrix, including such as is needed for
partial quenching.
 
However, for the calculation of the chiral condensate in this paper, 
we only need to check in Eq.~(\ref{eq:1-loop-calculation})
that the $U$ integral keeps the second term of NLO.
Since $\langle \xi^2_{ij}\rangle = \delta_{ij}\langle \xi^2_{ii}\rangle$, 
it is in fact enough to confirm that $[\mathcal{M}^\dagger(U-1)+ h.c.]_{ii}
\sim {\cal O}(p^3)$ or 
\begin{eqnarray} 
m_i  
\left(\frac{1}{2}\langle U_{ii}+U^\dagger_{ii}\rangle_U - 1\right)
\sim {\cal O}(p^3),
\end{eqnarray}
for any $i$, which can be done directly by means of the exact
group integration Eq.~(\ref{eq:cond}). Without using the rather complicated
exact expression, its asymptotic behavior is known 
\cite{Damgaard:2000di} for $m_i\Sigma V \ll 1$ and  $m_i\Sigma V \gg 1$,
and this leads to
\begin{eqnarray} 
m_i  
\left(\frac{1}{2}\langle U_{ii}+U^\dagger_{ii}\rangle_U - 1\right)
\to \left\{
\begin{array}{cc}
\frac{\nu}{\Sigma V} & (m_i\Sigma V \to 0)\\
-  \left(N_f-\frac{n_i}{2}\right) \frac{1}{\Sigma V} & (m_i\Sigma V \to \infty)
\end{array}
,\right.
\end{eqnarray}
with the other masses $m_j$'s ($j\neq i$) fixed, 
where $n_i$ denotes the degeneracy of the mass $m_i$ 
(Note that $n_i=0$ in the partially quenched case).
Since this function is everywhere regular for finite  
$m_i \Sigma V$, one expects that the two limiting cases above
are smoothly connected and the function thus always 
kept small, here of ${\cal O}(1/V)\sim {\cal O}(p^4)$
\footnote{We note that $\mathcal{M}(U-1)$ contributes
not as ${\cal O}(p^3)$ but even further suppressed, of ${\cal O}(p^4)$.
This is due to the fact that we are here considering a one-point
function.}.

In Fig.\ref{fig:MU-M},
we plot the function $m_i \Sigma  
\left(\langle U_{ii}+U^\dagger_{ii}\rangle_U/2 - 1\right)$
for various cases in a $(2+1)$-flavor theory.
Every curve shows a monotonous function connecting the 
two limits, thus confirming that the
$\mathcal{M}(U-1)\xi^2$ 
contribution to the condensate is always of order $1/V$. We provide
an explicit analytical expression in an analogous
$U(1)$ toy model in Appendix \ref{app:zero-mode-first}.

\FIGURE[tbp]{
\epsfig{file=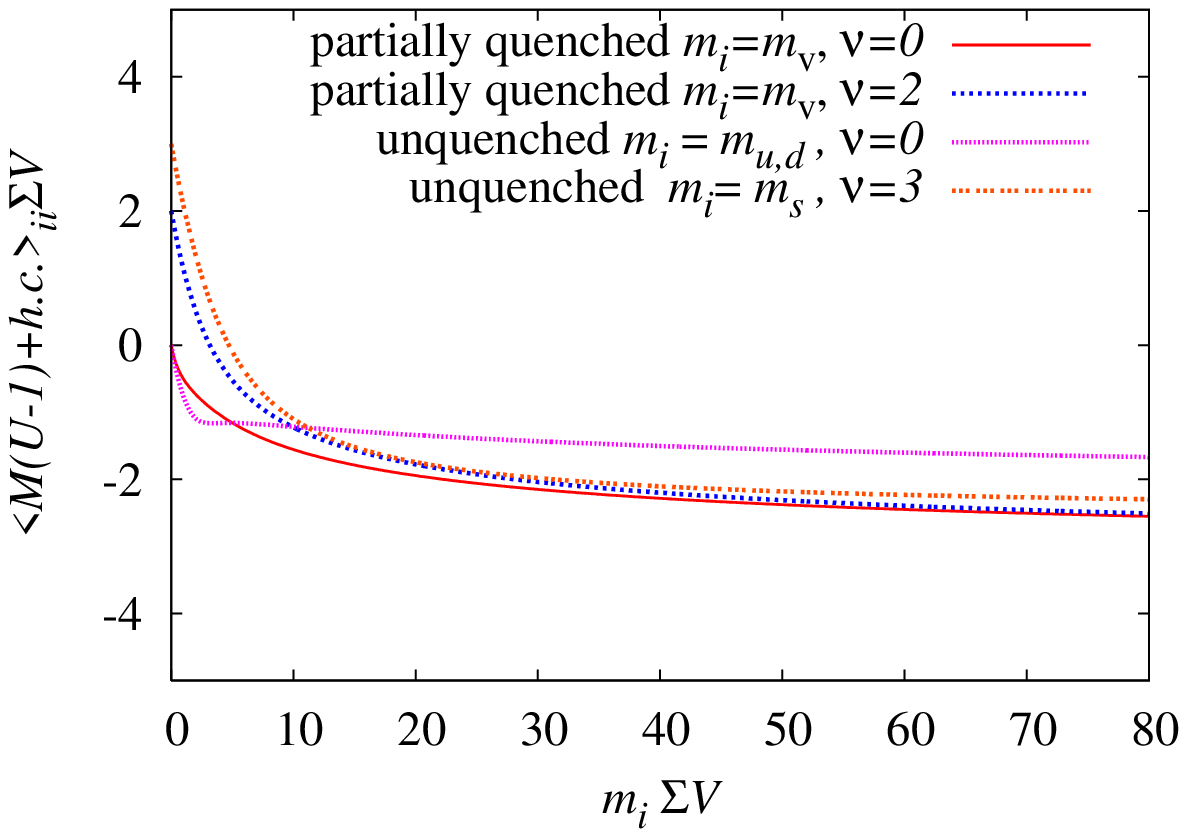, width=10cm}
  \caption{
The function $m_i \Sigma V 
\left(\langle U_{ii}+U^\dagger_{ii}\rangle_U/2 - 1\right)$
is plotted in various cases in 2+1 flavor theory:
 A. partially quenched theory ($m_i=m_v$) with 
$m_{u,d}\Sigma V = 5, m_s \Sigma V= 40$ at $\nu=0$ ($m_{u,d,s}$
denotes the quark mass of $u,d,s$ flavors respectively.),
 B. The same as A but at $\nu=2$, 
C.  unquenched theory with $m_i=m_{u,d}$ at $\nu=0$ where $m_s \Sigma V=40$ is fixed.
 D.  unquenched theory with $m_i=m_{s}$ at $\nu=3$ where $m_{u,d} \Sigma V=5$ 
   is fixed.
Every curve is kept at ${\cal O}(1)$ at any value of $m_i$,
which confirms $\mathcal{M}(U-1)$ always can be considered as NLO
to the calculation of the chiral condensate.
} 
  \label{fig:MU-M}
}
\section{A few examples}
\label{sec:examples}

In this section, we present two explicit numerical examples.
One is an $N_f=2$ degenerate two-flavor theory and the 
other is an $N_f=2+1$ theory including a strange quark
whose mass is different from up and down quark masses.
For the low-energy constants in both cases, 
we take the following phenomenologically reasonable values:
$\Sigma^{1/3}=250$ MeV, $F=90$ MeV, 
$L^r_6(\mu_{sub}=0.77{\rm GeV})=0.05\times 10^{-3}$ and
$L^r_8(\mu_{sub}=0.77{\rm GeV})=0.5\times 10^{-3}$ where 
$\mu_{sub}$ denotes the subtraction scale.


\subsection{degenerate $N_f=2$}
Let us first consider the two-flavor theory
with degenerate up and down quark masses $m_u=m_d=m$.
The factor $Z_v=Z^0_v+\delta Z_v(m_v)$ is then explicitly given by
\begin{eqnarray}
Z^0_v &=& 1-\frac{1}{F^2}\left[
\frac{M_\pi^2}{16\pi^2}\ln \frac{M_\pi^2}{2\mu_{sub}^2}
+2\bar{g}_1\left(M_\pi^2/2\right)
\right.\nonumber\\&&\left.
+\frac{1}{2}\left\{
\frac{\beta_1}{\sqrt{V}}-M_\pi^2
\left(\frac{1}{16\pi^2}\ln V^{1/2}\mu_{sub}^2+\beta_2 \right)
\right\}
-32L^r_6(\mu_{sub})M_\pi^2
\right],
\end{eqnarray}
and
\begin{eqnarray}
\delta Z_v(m_v) &=& 
-\frac{1}{F^2}\left[
\frac{M_\pi^2}{16\pi^2}\left(1+\frac{m_v}{m}\right)
\ln \left(1+\frac{m_v}{m}\right)
+\frac{M_\pi^2}{16\pi^2}\frac{m_v}{m}\ln \frac{M_\pi^2}{2\mu_{sub}^2}
\right.\nonumber\\&&\left.
+2\bar{g}_1\left(\frac{M_\pi^2}{2}\left(1+\frac{m_v}{m}\right)\right)
-2\bar{g}_1\left(\frac{M_\pi^2}{2}\right)
-\frac{1}{2}\left\{
\bar{g}_1\left(M_\pi^2 \frac{m_v}{m} \right)
+\frac{\beta_1}{\sqrt{V}}
\right.\right.\nonumber\\&&\left.\left.
-M_\pi^2\left(
\frac{1}{16\pi^2}(\ln M_\pi^2\frac{m_v}{m}V^{1/2} +1 )
-\bar{g}_2\left(M_\pi^2 \frac{m_v}{m} \right)+\beta_2\right)
\right.\right.\nonumber\\&&\left.\left.
+M_\pi^2 \frac{m_v}{m}
\left(\frac{1}{16\pi^2}-\bar{g}_2
\left(M_\pi^2 \frac{m_v}{m} \right)+\frac{1}{8\pi^2}
\ln \left(\frac{M_\pi^2 m_v}{\mu_{sub}^2m}\right)\right)
\right\}
\right.\nonumber\\&&\left.
\hspace{0.5in}-4(2L^r_8(\mu_{sub})+H^r_2(\mu_{sub}))M_\pi^2 \frac{m_v}{m} 
\right],\nonumber\\
\end{eqnarray}
where $M_\pi^2=(m_u+m_d)\Sigma /F^2 =2m\Sigma/F^2$.
For the numerical implementation of $\bar{g_1}$ and $\bar{g}_2$,
see Appendix \ref{app:g1g2}.
Note that $\delta Z_v(m_v=0)=0$.

For the spectral density, we also need
\begin{eqnarray}
{\rm Re}\delta Z_v(i\lambda) &=& 
-\frac{1}{F^2}\left[
\frac{M_\pi^2}{16\pi^2}
\left(\frac{1}{2}\ln \left(1+\frac{\lambda^2}{m^2}\right)
-\frac{\lambda}{m}\arctan \frac{\lambda}{m}\right)
\right.\nonumber\\&&\left.
+2{\rm Re}\bar{g}_1\left(\frac{M_\pi^2}{2}
\left(1+\frac{i\lambda}{m}\right)\right)
-2\bar{g}_1\left(\frac{M_\pi^2}{2}\right)
-\frac{1}{2}\left\{
{\rm Re}\bar{g}_1\left(M_\pi^2 \frac{i\lambda}{m} \right)
+\frac{\beta_1}{\sqrt{V}}
\right.\right.\nonumber\\&&\left.\left.
-M_\pi^2\left(
\frac{1}{16\pi^2}(\ln M_\pi^2\frac{\lambda}{m}V^{1/2} +1 )
-{\rm Re}\bar{g}_2\left(M_\pi^2 \frac{i\lambda}{m} \right)+\beta_2\right)
\right.\right.\nonumber\\&&\left.\left.
+M_\pi^2 \frac{\lambda}{m}\left({\rm Im} \bar{g}_2
\left(M_\pi^2 \frac{i\lambda}{m} \right)-\frac{1}{16\pi}\right)
\right\}\right].
\end{eqnarray}
As discussed above, the $H_2$ (and $\mu_{sub}$) dependence has disappeared.

The non-perturbative expressions for the zero-mode 
integrals are given by \cite{Splittorff:2002eb}
\begin{eqnarray}
\label{eq:Sigma-2flavor}
\hat{\Sigma}^{\rm PQ}_\nu (\mu_v,\mu)
&=&
-\frac{1}{(\mu^{2}-\mu_v^2)^2}
\nonumber\\
&&\hspace{-1in}
\times \frac{\det \left(
\begin{array}{cccc}
\partial_{\mu_v}K_\nu(\mu_{v}) & I_\nu(\mu_v) & I_\nu(\mu) & 
\mu^{-1}I_{\nu-1}(\mu) \\
-\partial_{\mu_v}(\mu_{v} K_{\nu+1}(\mu_{v})) 
& \mu_v I_{\nu+1}(\mu_v) & \mu I_{\nu+1}(\mu) & I_{\nu}(\mu)\\
\partial_{\mu_v}(\mu_{v}^2 K_{\nu+2}(\mu_{v})) & \mu_v^2I_{\nu+2}(\mu_v) & 
\mu^2I_{\nu+2}(\mu) 
& \mu I_{\nu+1}(\mu) \\
-\partial_{\mu_v}(\mu_{v}^3 K_{\nu+3}(\mu_{v})) & \mu_v^3I_{\nu+3}(\mu_v) & 
\mu^3I_{\nu+3}(\mu) 
& \mu^2 I_{\nu+2}(\mu) \\
\end{array}
\right)}{\det \left(
\begin{array}{cc}
I_\nu(\mu) & \mu^{-1}I_{\nu-1}(\mu) \\
\mu I_{\nu+1}(\mu) & I_{\nu}(\mu)\\
\end{array}
\right)},\nonumber\\
\end{eqnarray}
and
\begin{eqnarray}
\label{eq:rho-2flavor}
\hat{\rho}^{mic}_\nu (\zeta,\mu)
&=&
\frac{|\zeta|}{2(\mu^{2}+\zeta^2)^2}
\nonumber\\
&&\hspace{-1in}
\times \frac{\det \left(
\begin{array}{cccc}
\zeta^{-1}J_{\nu-1}(\zeta) & J_\nu(\zeta) & I_\nu(\mu) & \mu^{-1}I_{\nu-1}(\mu) \\
J_{\nu}(\zeta) & \zeta J_{\nu+1}(\zeta) & -\mu I_{\nu+1}(\mu) & -I_{\nu}(\mu)\\
\zeta J_{\nu+1}(\zeta) & \zeta^2 J_{\nu+2}(\zeta) &\mu^2I_{\nu+2}(\mu) & \mu I_{\nu+1}(\mu) \\
\zeta^2J_{\nu+2}(\zeta)& \zeta^3J_{\nu+3}(\zeta) & -\mu^3I_{\nu+3}(\mu) & -\mu^2 I_{\nu+2}(\mu) \\
\end{array}
\right)}{\det \left(
\begin{array}{cc}
I_\nu(\mu) & \mu^{-1}I_{\nu-1}(\mu) \\
\mu I_{\nu+1}(\mu) & I_{\nu}(\mu)\\
\end{array}
\right)},\nonumber\\
\end{eqnarray}
where $\mu_v=m_v\Sigma V$, $\mu=m \Sigma V$ and $\zeta=\lambda \Sigma V$.\\

Substituting the explicit expressions above into Eq.~(\ref{eq:condNLO-expand})
and (\ref{eq:general-rho_nu2}),
it is straightforward to compute numerically 
the chiral condensate and the spectral density of the Dirac operator.
Using the numerical values listed in the beginning of this section,
we plot the chiral condensate and the spectral density in 
Fig.~\ref{fig:Nf2-3regimes}
for the case of $m_u=m_d=$ 10 MeV, $L=T/2=2$ fm, $\nu=0$ and
here taking just as an example
$H^r_2(0.77{\rm GeV})=0.1\times 10^{-3}$.
One sees that the resulting thick line 
smoothly connects the one in the $\epsilon$-regime (solid)
with the one in the 
$p$-regime (dotted)\footnote{Because we do not have $p$-regime
predictions available at fixed topological charge $\nu$, we always compare
with $p$-regime expressions where topological sectors have been summed over.
Of course, $p$-regime predictions at fixed topology will differ slightly from
these, but the difference becomes insignificant at large volumes.}.
It is clear that the effect of one-loop corrections to 
the spectral density of the Dirac operator in the
$p$-expansion is relatively small
when $N_f=2$, in agreement with the prediction 
Eq.~(\ref{eq:Smilga-Stern}) of Smilga and Stern \cite{Smilga:1993in}.\\

It is useful to see what happens when we vary some parameters.
In the larger volume ($L=T/2=3$ fm) shown in Fig.~\ref{fig:large}
the agreement between the three different expansions becomes visibly better
around $ML\sim $2-3. 
The sea quark mass dependence is shown
in Fig.~\ref{fig:msea} where we show
the plots for different sea quark masses at 2 MeV (solid), 
10 MeV (dotted), and 30 MeV (small dotted).
The black filled circle in the left part shows the physical 
points where $m_v=m_u=m_d$. 
In Fig.~\ref{fig:nu} the $\nu$-dependence is presented 
at a fixed value of $m_u=10$ MeV and $L=T/2=$ 2fm.
The topology dependence disappears around $M_{vv}L=3$-4.
The condensate of the physical theory with $m_v=m_u=m_d$ 
is plotted on the left part of Fig.~\ref{fig:theta}
after summing of topology 
(See Appendix \ref{app:zero-mode} for the details).
Of course, the condensate in the massive case is inherently
ambiguous due to the presence of the coupling $H_2$.
On the other hand, the spectral density is free from this ambiguity.
On the right part of Fig.~\ref{fig:theta} we plot the
density after summing over topology at fixed $m_u=m_d=10$ MeV.
As has been observed before \cite{Damgaard:1999ij}, 
the striking oscillations that are seen at sectors of
fixed topological charge become
smeared out upon summing over topology. In
the massless limit the oscillations will reappear and the density
approaches that of the $\nu = 0$ sector.\\

\FIGURE[t]{
\epsfig{file=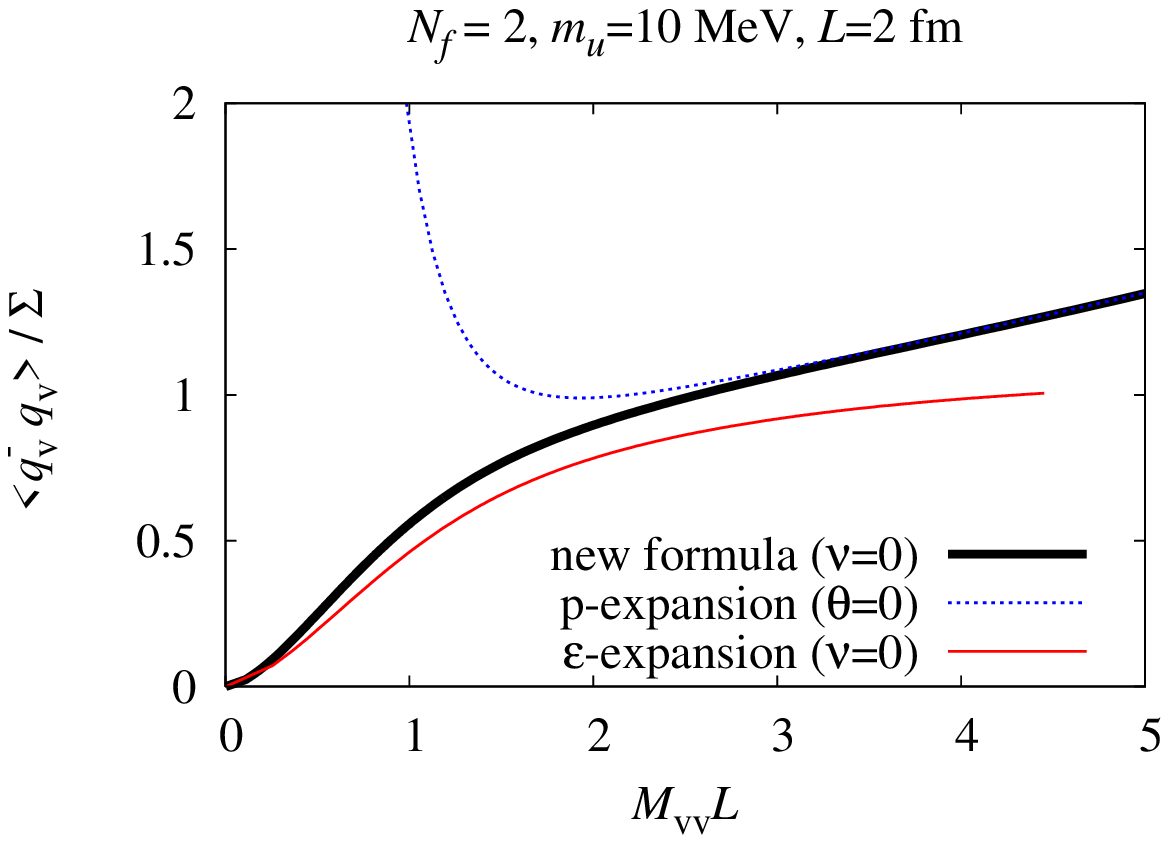,width=12.5cm}
\epsfig{file=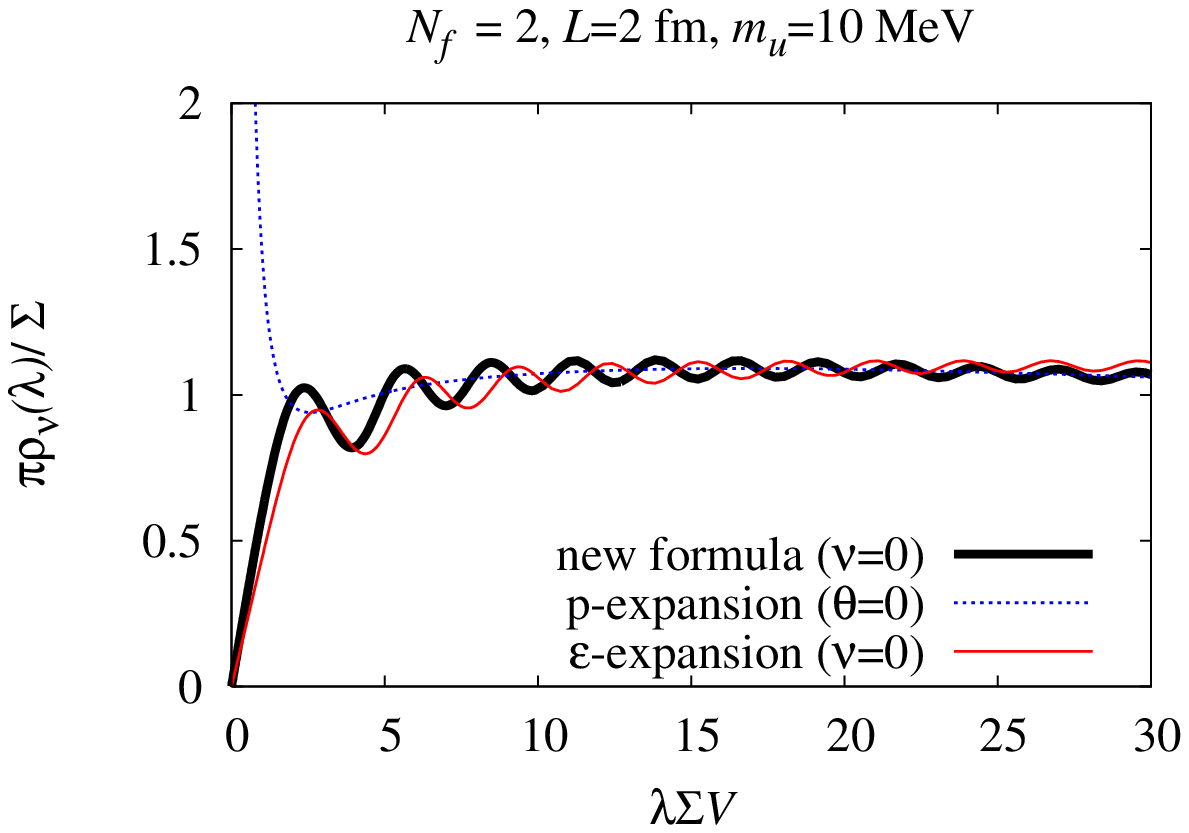,width=12.5cm}
  \caption{
The chiral condensate (top) and Dirac spectral density (bottom)
for the case with $m_u=m_d=10$ MeV, $L=T/2=2$ fm.
The result of this paper is given by the thick line which smoothly 
connects the one in the $\epsilon$-regime (solid)
with the on of $p$-regime (dotted).
Parameters have been chosen as
$\Sigma^{1/3}=250$ MeV, $F=90$ MeV, 
$L^r_6(\mu_{sub}=0.77{\rm GeV})=0.05\times 10^{-3}$,
$L^r_8(\mu_{sub}=0.77{\rm GeV})=0.5\times 10^{-3}$
and $H^r_2(\mu_{sub}=0.77{\rm GeV})=0.1\times 10^{-3}$.
}
  \label{fig:Nf2-3regimes}
}

\FIGURE[tp]{
\epsfig{file=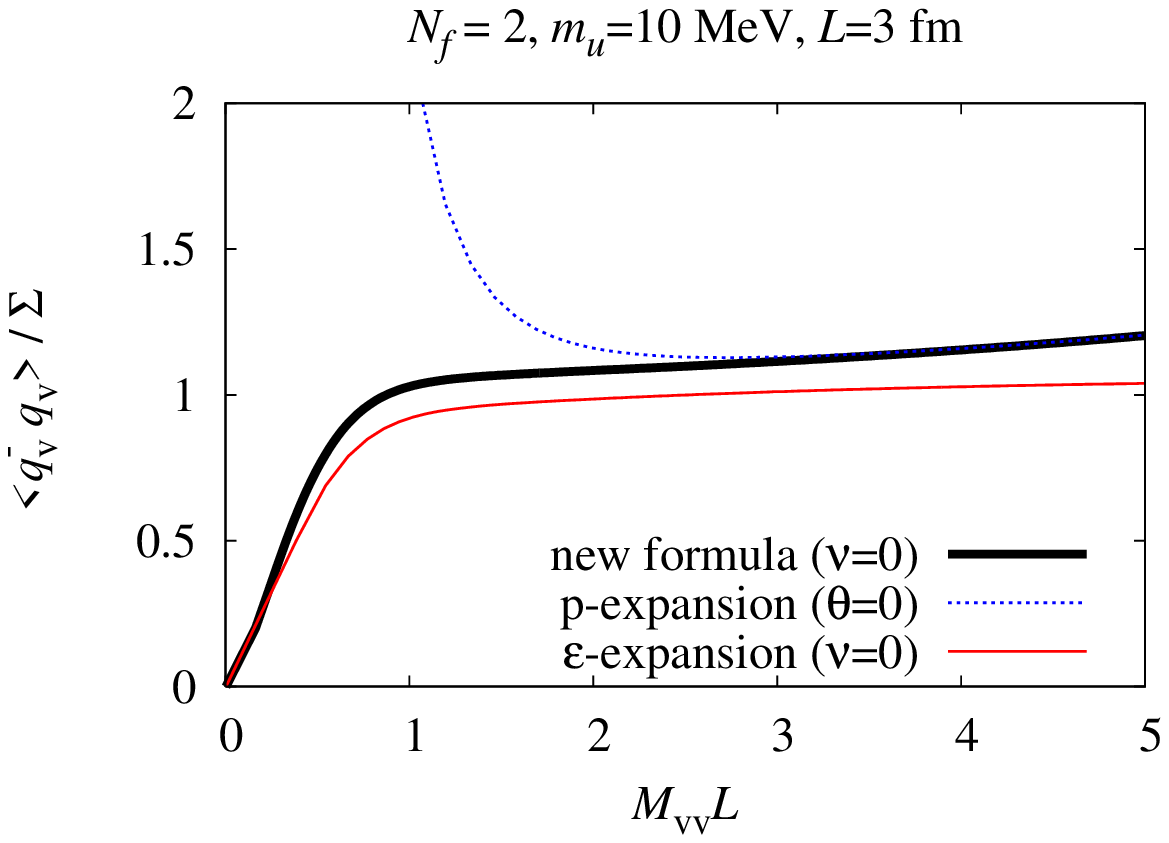,width=7.4cm}
\epsfig{file=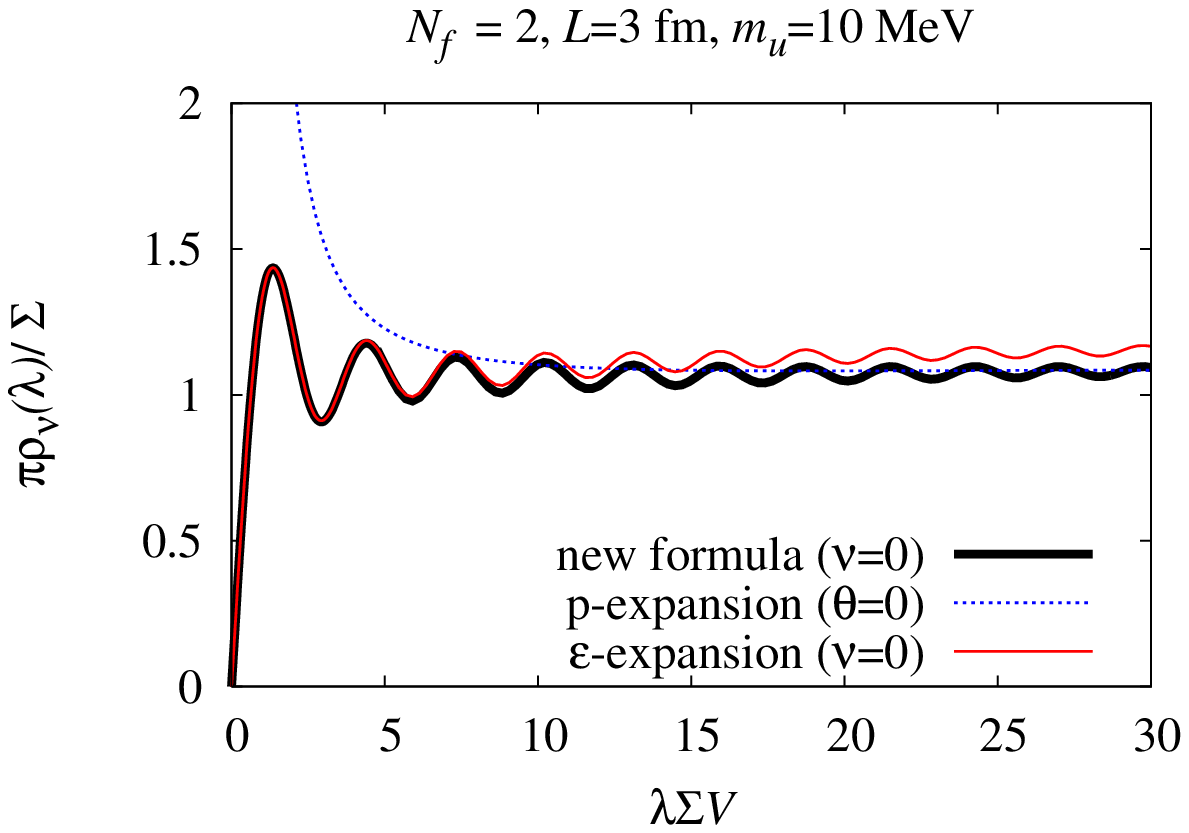,width=7.4cm}
  \caption{
The same plots as in Fig.~\ref{fig:Nf2-3regimes},
the condensate (left) and the spectral density (right)
but with the larger volume size
$L=T/2=3$ fm. The convergence of three expansions 
around $M_{vv}L\sim 2$-3 or $\lambda \Sigma V\sim$ 10
becomes better. 
}
  \label{fig:large}
}
\FIGURE[tp]{
\epsfig{file=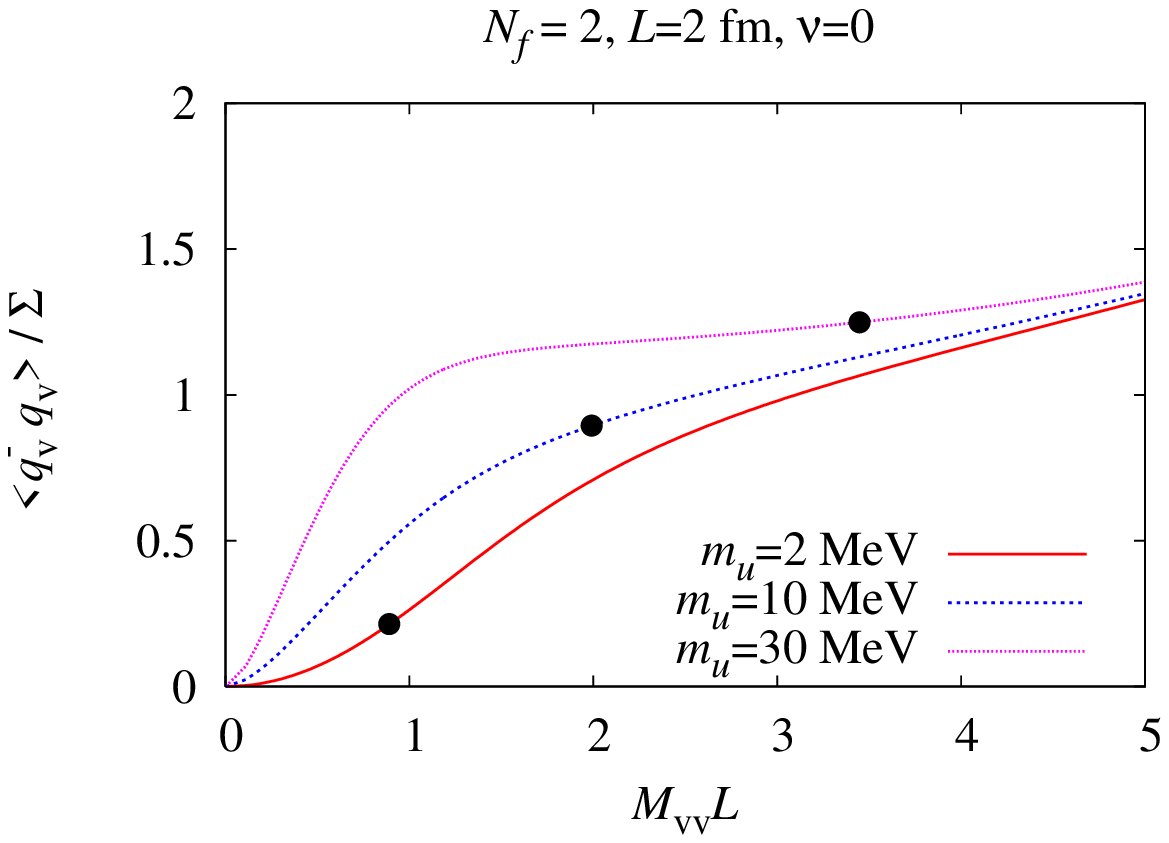,width=7.4cm}
\epsfig{file=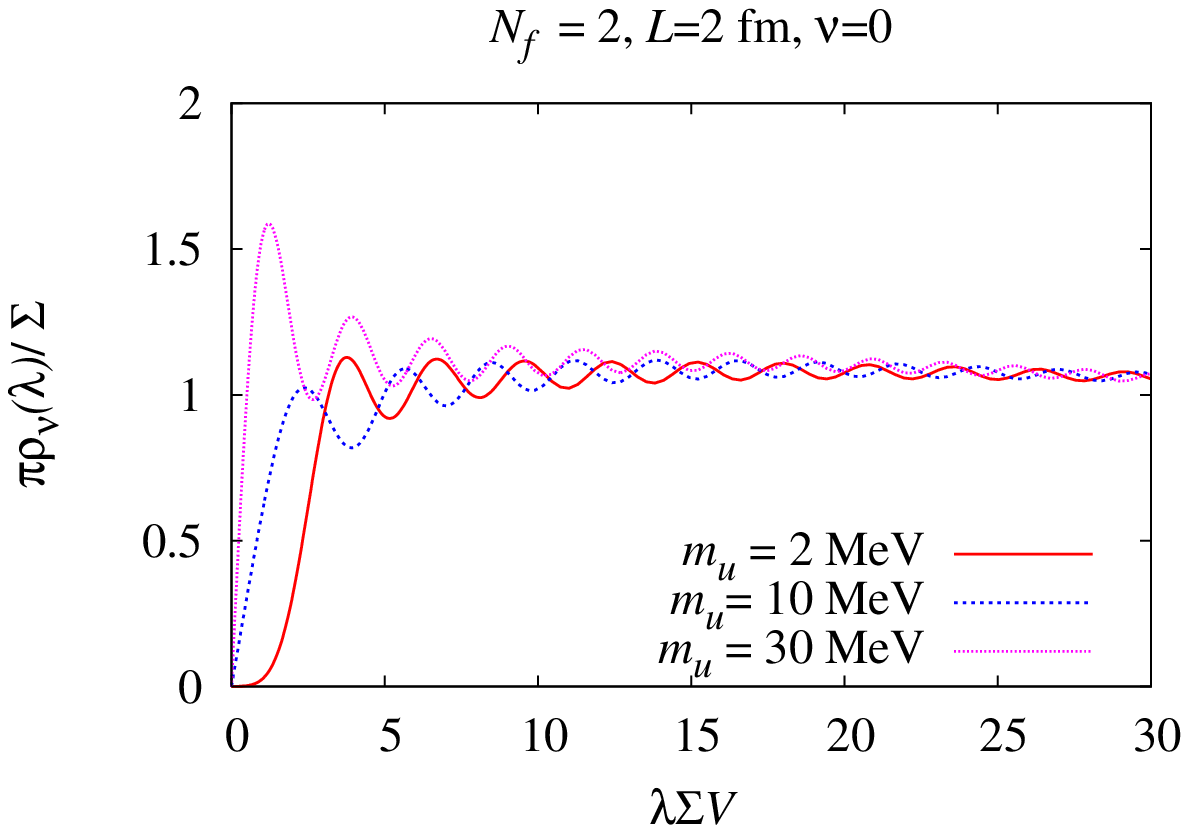,width=7.4cm}
  \caption{
The condensate (left) and the spectral density (right)
at fixed topology $\nu=0$ 
for different sea quark masses at 2 MeV (solid), 
10 MeV (dotted), 30 MeV (small dotted) are given.
We again set $L=T/2=$ 2fm.
The black filled circle in the left panel shows the physical 
points where $m_v=m_u=m_d$. 
}
  \label{fig:msea}
}
\FIGURE[t]{
\epsfig{file=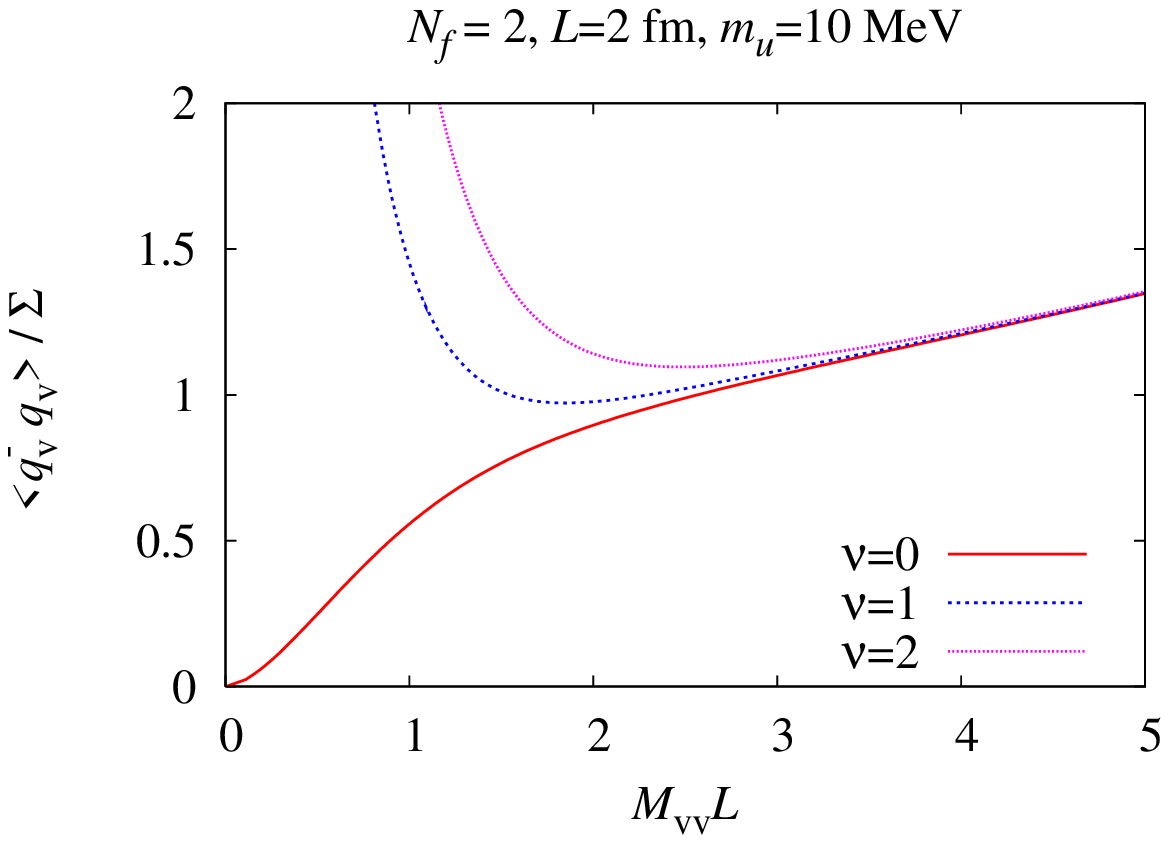,width=7.4cm}
\epsfig{file=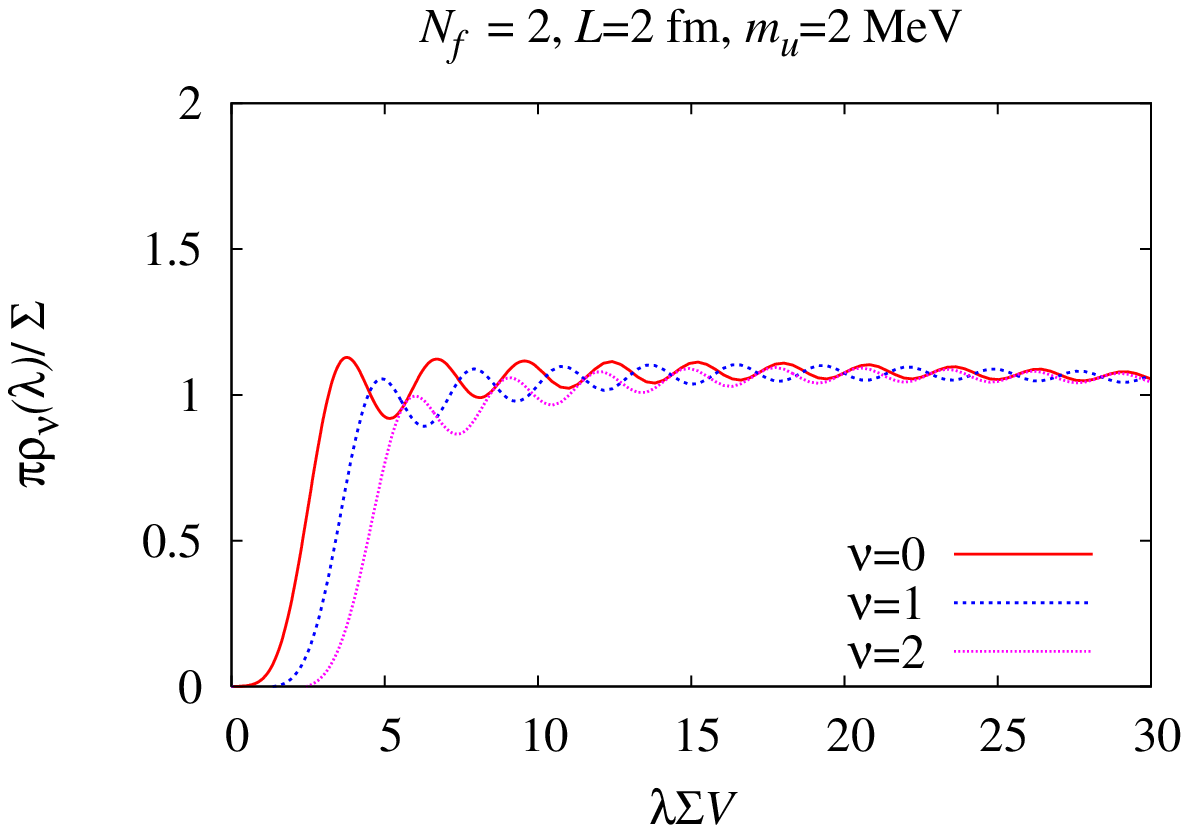,width=7.4cm}
  \caption{
The topology $\nu$ dependence of 
the condensate (left) and the spectral density (right)
at a fixed value of $m_u=10$ MeV and $L=T/2=$ 2fm and
$H^r_2(0.77{\rm GeV})=0.1\times 10^{-3}$.
The topology dependence becomes negligible around $M_{vv}L=3$-4.
}
  \label{fig:nu}
}
\FIGURE[t]{
\epsfig{file=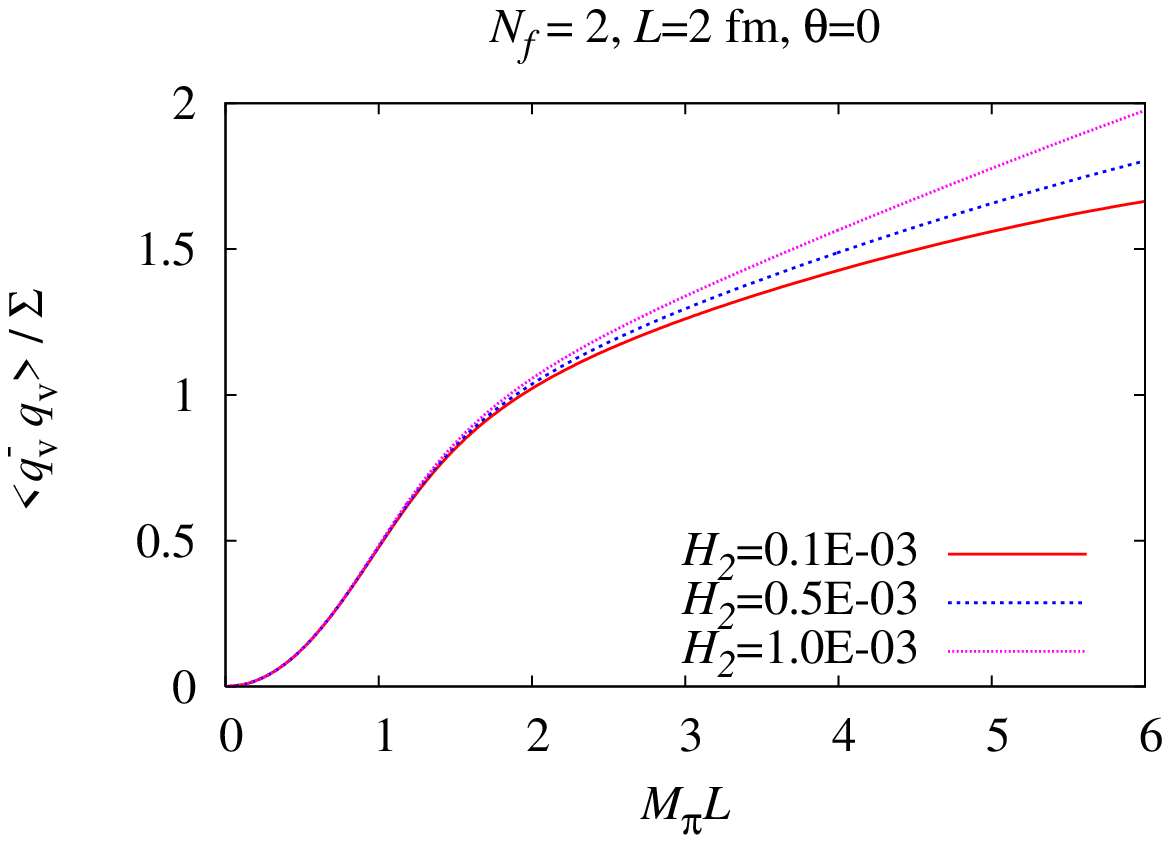,width=7.4cm}
\epsfig{file=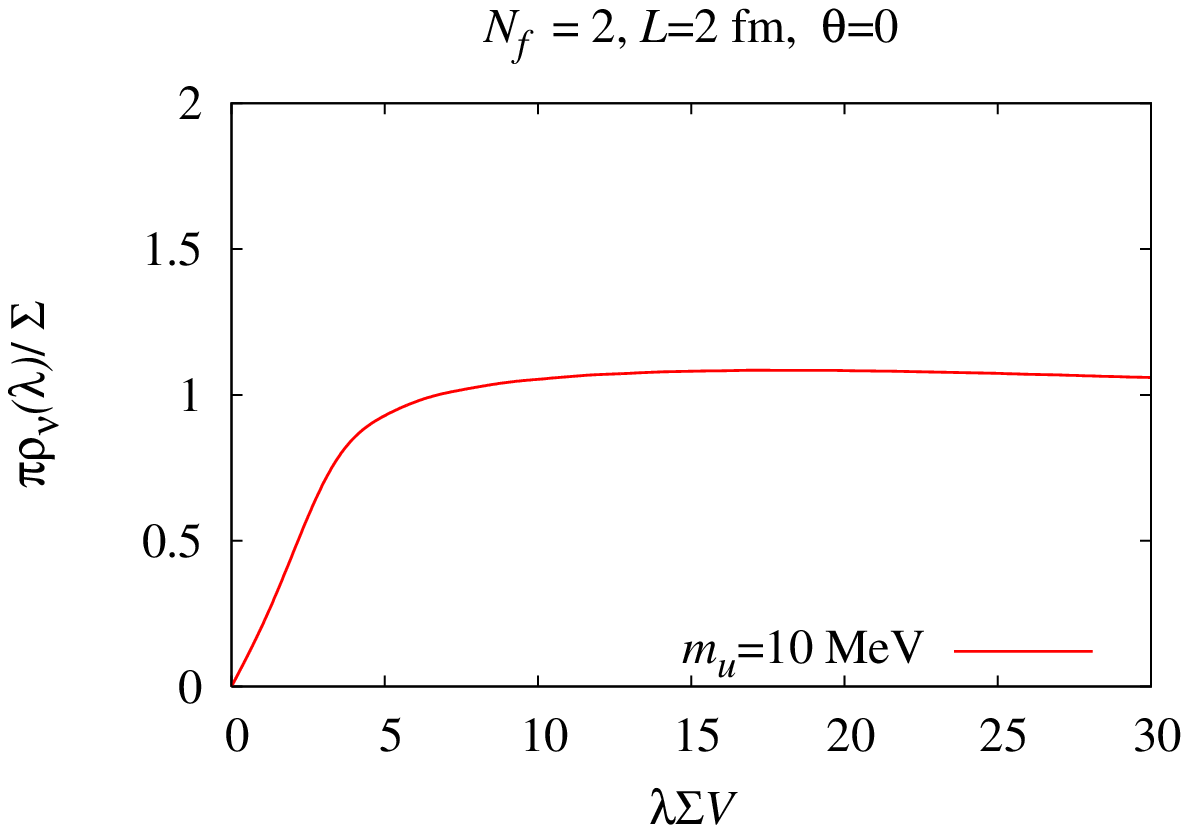,width=7.4cm}
  \caption{
The (full theory) condensate (left) where $m_v=m_u=m_d$
and the spectral density (right) at fixed $m_u=m_d=10$ MeV,
are given at $\theta=0$.
Due to the ambiguity of $H_2^r$, the condensate
is not a well-defined quantity when $M_{\pi}$ or $M_{vv}\neq 0$,
while the spectral density is free from this ambiguity.
}
  \label{fig:theta}
}

\subsection{$N_f=2+1$}

Next, we consider the $N_f=2+1$ theory where the strange quark mass
is different from the up and down quark masses.
Then $\bar{G}(x,M^2,M^2)$ becomes much more complicated,
and we refer to the
explicit expressions given in Eqs. (\ref{eq:Gbar2+1})
and (\ref{eq:ABC2+1}).

One obtains
\begin{eqnarray}
Z^0_v &=& 
1-\frac{1}{F^2}\left[
\frac{M_{ud}^2}{16\pi^2}\ln \frac{M_{ud}^2}{2\mu_{sub}^2}
+2\bar{g}_1\left(M_{ud}^2/2\right)
+\frac{M_{ss}^2}{32\pi^2}\ln \frac{M_{ss}^2}{2\mu_{sub}^2}
+\bar{g}_1\left(M_{ss}^2/2\right)
\right.\nonumber\\&&\left.
-\frac{1}{3}\left\{
-\frac{2(M_{ud}^2-M_{ss}^2)^2}{(M_{ud}^2+2M_{ss}^2)^2}
\left(\frac{M_{ud}^2+2M_{ss}^2}{3\times 16\pi^2}
\ln \frac{M_{ud}^2+2M_{ss}^2}{3\mu_{sub}^2}
+\bar{g}_1\left(\frac{M_{ud}^2+2M_{ss}^2}{3}\right)\right)
\right.\right.\nonumber\\&&\left.\left.
+\left(1+\frac{2(M_{ud}^2-M_{ss}^2)^2}{(M_{ud}^2+2M_{ss}^2)^2}\right)
\left(-\frac{\beta_1}{\sqrt{V}}\right)
-\frac{3M_{ud}^2M_{ss}^2}{M_{ud}^2+2M_{ss}^2}
\left(-\frac{1}{16\pi^2}\ln V^{1/2}\mu_{sub}^2-\beta_2 \right)
\right\}
\right.\nonumber\\&&\left.
-16L^r_6(\mu_{sub})(2M_{ud}^2 + M_{ss}^2)\right],
\end{eqnarray}
and 
\begin{eqnarray}
\delta Z_v(m_v) &=& 
-\frac{1}{F^2}\left[
\frac{M_{ud}^2}{16\pi^2}\left(1+\frac{m_v}{m_u}\right)
\ln \left(1+\frac{m_v}{m_u}\right)
+\frac{M_{ud}^2}{16\pi^2}\frac{m_v}{m_u}\ln \frac{M_{ud}^2}{2\mu_{sub}^2}
\right.\nonumber\\&&\left.
+2\bar{g}_1\left(\frac{M_{ud}^2}{2}\left(1+\frac{m_v}{m_u}\right)\right)
-2\bar{g}_1\left(\frac{M_{ud}^2}{2}\right)
+\frac{M_{ss}^2}{32\pi^2}\left(1+\frac{m_v}{m_s}\right)
\ln \left(1+\frac{m_v}{m_s}\right)
\right.\nonumber\\&&\left.
+\frac{M_{ss}^2}{32\pi^2}\frac{m_v}{m_s}\ln \frac{M_{ss}^2}{2\mu_{sub}^2}
+\bar{g}_1\left(\frac{M_{ss}^2}{2}\left(1+\frac{m_v}{m_s}\right)\right)
-\bar{g}_1\left(\frac{M_{ss}^2}{2}\right)
\right.\nonumber\\&&\left.\hspace{-.5in}
-\frac{1}{3}\left\{
-\frac{2(M_{ud}^2-M_{ss}^2)^2}{(3M_{vv}^2-M_{ud}^2-2M_{ss}^2)^2}
\left(\frac{M_{ud}^2+2M_{ss}^2}{3\times 16\pi^2}
\ln \frac{M_{ud}^2+2M_{ss}^2}{3\mu_{sub}^2}
+\bar{g}_1\left(\frac{M_{ud}^2+2M_{ss}^2}{3}\right)\right)
\right.\right.\nonumber\\&&\left.\left.
+\left(1+\frac{2(M_{ud}^2-M_{ss}^2)^2}{(3M_{vv}^2-M_{ud}^2-2M_{ss}^2)^2}\right)
\left(\frac{M_{vv}^2}{16\pi^2}\ln \frac{M_{vv}^2}{\mu_{sub}^2}
+\bar{g}_1(M_{vv}^2)\right)
\right.\right.\nonumber\\&&\left.\left.
+\frac{3(M_{vv}^2-M_{ud}^2)(M_{vv}^2-M_{ss}^2)}{3M_{vv}^2-M_{ud}^2-2M_{ss}^2}
\left(\frac{1}{16\pi^2}(\ln \frac{M_{vv}^2}{\mu_{sub}^2}+1) 
+\partial_{M^2}\bar{g}_1(M_{vv}^2)  \right)
\right.\right.\nonumber\\&&\left.\left.
+\frac{2(M_{ud}^2-M_{ss}^2)^2}{(M_{ud}^2+2M_{ss}^2)^2}
\left(\frac{M_{ud}^2+2M_{ss}^2}{3\times 16\pi^2}
\ln \frac{M_{ud}^2+2M_{ss}^2}{3\mu_{sub}^2}
+\bar{g}_1\left(\frac{M_{ud}^2+2M_{ss}^2}{3}\right)\right)
\right.\right.\nonumber\\&&\left.\left.\hspace{-0.5in}
-\left(1+\frac{2(M_{ud}^2-M_{ss}^2)^2}{(M_{ud}^2+2M_{ss}^2)^2}\right)
\left(-\frac{\beta_1}{\sqrt{V}}\right)
+\frac{3M_{ud}^2M_{ss}^2}{M_{ud}^2+2M_{ss}^2}
\left(-\frac{1}{16\pi^2}\ln V^{1/2}\mu_{sub}^2-\beta_2 \right)\right\}
\right.\nonumber\\ && \left.
-4(2L_8^r(\mu_{sub}) + H_2^r(\mu_{sub}))M_{vv}^2\right],
\end{eqnarray}
where its real part for the imaginary valence mass
is taken as in the same way shown in the previous 
subsection.
Here $M_{ud}^2=(m_u+m_d)\Sigma/F^2, M_{ss}^2=2m_s \Sigma /F^2, 
M_{vv}^2=2m_v \Sigma /F^2$.

The non-perturbative zero-mode integrals are given by
\begin{eqnarray}
\label{eq:Sigma-3flavor}
\hat{\Sigma}^{\rm PQ}_\nu (\mu_v,\{\mu_u, \mu_s\})
&=& -\frac{1}{(\mu_{u}^{2}-\mu_v^2)^2(\mu_s^2-\mu_v^2)}
\nonumber\\
&&\hspace{-1in}
\times \frac{\det \left(
\begin{array}{ccccc}
\partial_{\mu_v}K_\nu(\mu_v) & I_\nu(\mu_v) & I_\nu(\mu_{u}) & 
\mu_u^{-1}I_{\nu-1}(\mu_{u}) & I_\nu(\mu_{s})\\
-\partial_{\mu_v} (\mu_{v} K_{\nu+1}(\mu_{v})) & \mu_v I_{\nu+1}(\mu_v) 
& \mu_{u}I_{\nu+1}(\mu_{u}) 
& I_{\nu}(\mu_{u}) &  \mu_{s}I_{\nu+1}(\mu_{s}) \\
\partial_{\mu_v} (\mu_{v}^2 K_{\nu+2}(\mu_{v})) & \mu_v^2I_{\nu+2}(\mu_v) & 
\mu_{u}^2I_{\nu+2}(\mu_{u}) 
& \mu_{u}I_{\nu+1}(\mu_{u}) & \mu_{s}^2I_{\nu+2}(\mu_{s})\\
-\partial_{\mu_v} (\mu_{v}^3 K_{\nu+3}(\mu_{v})) & \mu_v^3I_{\nu+3}(\mu_v) & 
\mu_{u}^3I_{\nu+3}(\mu_{u}) 
& \mu_{u}^2I_{\nu+2}(\mu_{u}) & \mu_{s}^3I_{\nu+3}(\mu_{s})\\
\partial_{\mu_v} (\mu_{v}^4 K_{\nu+4}(\mu_{v})) & \mu_v^4I_{\nu+4}(\mu_v) & 
\mu_{u}^4I_{\nu+4}(\mu_{u}) 
& \mu_{u}^3I_{\nu+3}(\mu_{u}) & \mu_{s}^4I_{\nu+4}(\mu_{s})
\end{array}
\right)}{\det \left(
\begin{array}{ccc}
I_\nu(\mu_{u}) & \mu_u^{-1}I_{\nu-1}(\mu_{u}) & I_\nu(\mu_{s})\\
\mu_{u}I_{\nu+1}(\mu_{u}) & I_{\nu}(\mu_{u}) &  \mu_{s}I_{\nu+1}(\mu_{s}) \\
\mu_{u}^2I_{\nu+2}(\mu_{u}) & \mu_{u}I_{\nu+1}(\mu_{u}) & \mu_{s}^2I_{\nu+2}(\mu_{s})\\
\end{array}
\right)}.\nonumber\\
\end{eqnarray}
where $\mu_u=m_{u,d}\Sigma V$ and $\mu_s=m_s\Sigma V$ and
one can take the imaginary value of $\mu_v=i \zeta$ to 
obtain $\hat{\rho}^{mic}_\nu(\zeta,\{\mu_u, \mu_s\})$ 
in the same way as the 2-flavor case.

Substituting again the numerical values in the beginning of this section,
we plot the chiral condensate and the spectral density in 
Fig.~\ref{fig:Nf3}
for the case with $m_u=m_d=$ 10 MeV, $m_s= 110$ MeV,
$L=T/2=2$ fm, $\nu=0$ and $H^r_2(0.77{\rm GeV})=0.1\times 10^{-3}$.
Due to the large contribution from the chiral logarithm
of strange quarks, 
both of the (normalized) condensate and spectral density shows a quite
larger deviation from 1, which is the universal chiral 
limit when $V\to \infty$ with any number of flavors.

\FIGURE[t]{
\epsfig{file=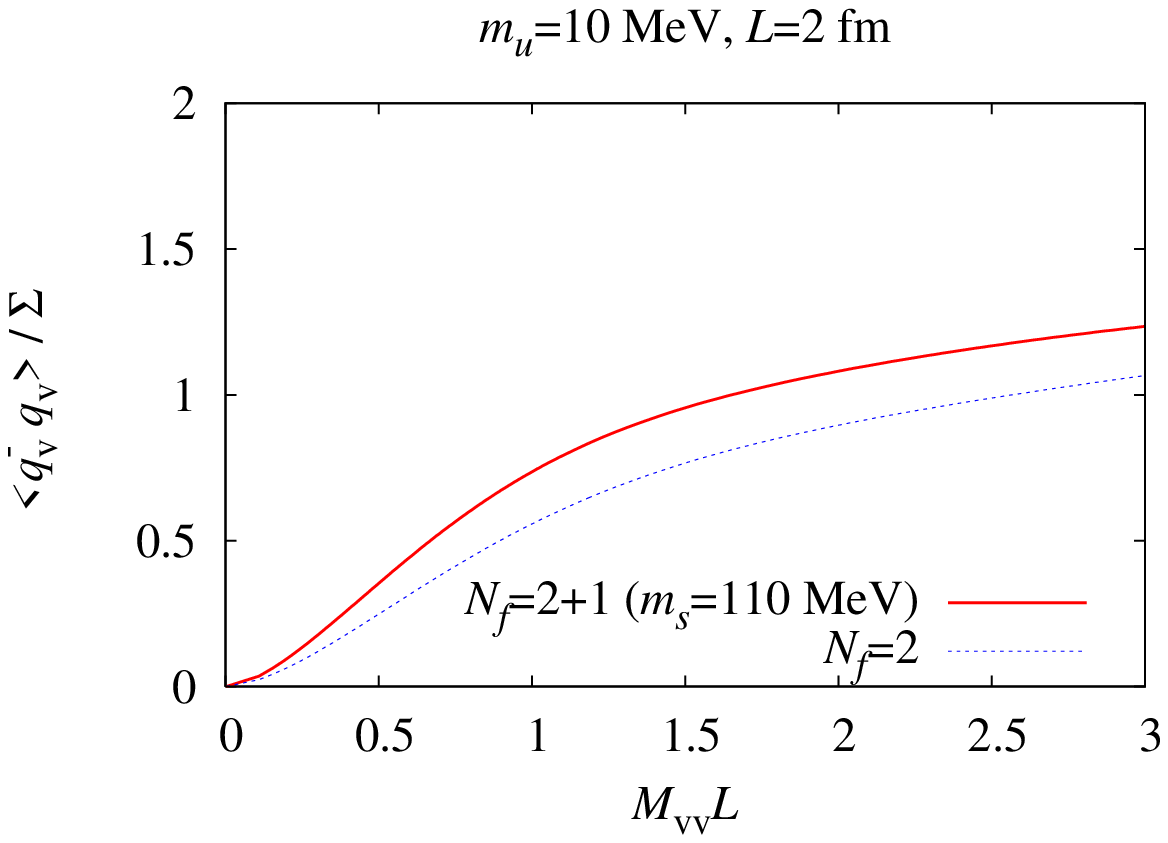,width=12.5cm}
\epsfig{file=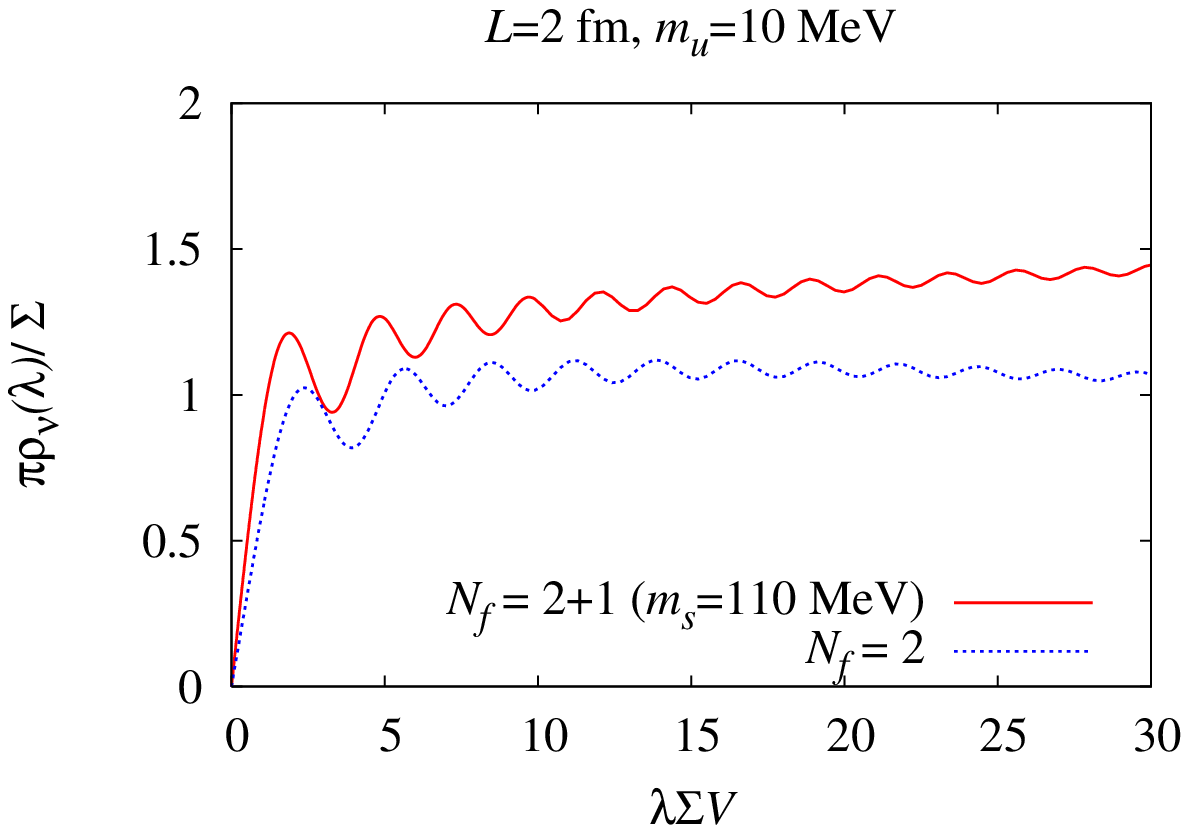,width=12.5cm}
  \caption{
The condensate (top) and the Dirac spectral density (bottom)
with $N_f=2+1$ theory are given (solid curve).
For the comparison, $N_f=2$ plot is also presented (dotted).
As the valence mass or the eigenvalue increases, 
both of them show larger deviation from 1 than the case with $N_f=2$.
}
  \label{fig:Nf3}
}

\section{Conclusions}
\label{sec:conclusion}

We have computed the 
quark condensate and the spectral density of the Dirac operator
by means of the chiral Lagrangian in a finite volume. 
We have used a new perturbative method which keeps 
all the terms which appear in the conventional expansion 
in the $p$-regime but which treats
zero-mode integral non-perturbatively  
in the same manner as in the $\epsilon$-regime.
The resulting perturbative series connects smoothly 
previous results in the $p$-regime 
with those of the $\epsilon$-regime. Having analytical
formulas available in the intermediate region may be of 
obvious value for lattice QCD simulations. \\


It would be interesting to investigate the present proposal
for finite-volume chiral perturbation theory in greater detail.
One should obviously consider using such a method 
to compute other
finite-volume observables in chiral perturbation theory, such
as two-point correlation functions.

\acknowledgments

We thank Fabio Bernardoni, Shoji Hashimoto, Pilar Hernandez, Martin
L\"{u}scher and Kim Splittorff for discussions.
The work of PHD was partly supported by the EU network
ENRAGE MRTN-CT-2004-005616.
The work of HF was supported by the 
Japanese Society for the Promotion of Science.

\appendix


\section{Non-perturbative zero-mode integrals}
\label{app:zero-mode}
\setcounter{equation}{0}

In this appendix, we briefly review how to 
perform the non-perturbative zero-mode integrals.
We are interested in the most general 
partially quenched calculations for both practical purposes
of comparisons to lattice data and for computing spectral
correlation functions of the Dirac operator. Of course,
results for the $N_f$-theory without separate valence quarks
are trivially included in this.

For the evaluation of the zero-mode integrals it is convenient
to use the graded formalism where partial quenching is achieved
by adding additional bosonic and fermionic species.
The zero-mode integral corresponding to the graded version of
Eq.~(\ref{Znuzero})  
with $n$ bosons and $m$ fermions is known 
in closed analytical form for an arbitrary mass matrix
\cite{Splittorff:2002eb},
\begin{equation}
\mathcal{Z}^\nu_{n,m}(\{\mu_i\})
=
\frac{\det[\mu_i^{j-1}\mathcal{J}_{\nu +j-1}(\mu_i)]_{i,j=1,\cdots n+m}}
{\prod_{j>i=1}^n(\mu_j^2-\mu_i^2)\prod_{j>i=n+1}^{n+m}(\mu_j^2-\mu_i^2)},
\end{equation}
where $\mu_i=m_i\Sigma V$.
Here $\mathcal{J}$'s are defined as
$\mathcal{J}_{\nu+j-1}(\mu_i)\equiv (-1)^{j-1} K_{\nu+j-1}(\mu_i)$ 
for $i=1,\cdots n$ and 
$\mathcal{J}_{\nu+j-1}(\mu_i)\equiv I_{\nu+j-1}(\mu_i)$ 
for $i=n+1,\cdots n+m$, 
where $I_\nu$ and $K_\nu$ are the modified Bessel functions.
In this paper, we need the case with $(n, m)=(1, N_f+1)$:
\begin{eqnarray}
\label{eq:zero-mode1}
\mathcal{Z}^\nu_{1,1+N_f}(\mu_b |\mu_v ,\{\mu_{sea}\})
&=&
\frac{1}{\prod_{i=1}^{N_f}(\mu_{i}^{2}-\mu_v^2)
\prod_{j>k}^{N_f}(\mu_{j}^{2}-\mu_{k}^{2})}
\nonumber\\
&&\hspace{-1in}
\times \det \left(
\begin{array}{ccccc}
K_\nu(\mu_{b}) & I_\nu(\mu_v) & I_\nu(\mu_{1}) & I_\nu(\mu_{2}) & \cdots\\
-\mu_{b} K_{\nu+1}(\mu_{b}) & \mu_v I_{\nu+1}(\mu_v) & \mu_{1}I_{\nu+1}(\mu_{1}) 
& \mu_{2}I_{\nu+1}(\mu_{2}) & \cdots\\
\mu_{b}^2 K_{\nu+2}(\mu_{b}) & \mu_v^2I_{\nu+2}(\mu_v) & 
\mu_{1}^2I_{\nu+2}(\mu_{1}) 
& \mu^2_{2}I_{\nu+2}(\mu_{2}) & \cdots\\
\cdots & \cdots & \cdots & \cdots & \cdots
\end{array}
\right).
\end{eqnarray}
Here $\mu_b=m_b\Sigma V$,  
$\mu_v=m_v\Sigma V$ and $\mu_{i}=m_{i}\Sigma V$, 
where $m_b$, $m_v$ and $m_{i}$ denote the masses of 
the valence bosons, the valence quarks, and the physical sea 
quarks respectively.
In the arguments, the set of sea flavors are denoted by
$\{\mu_{sea}\}=\{m_1 \Sigma V, m_2 \Sigma V,\cdots\}$.\\

When some of the quark masses are degenerate, one simply uses
l'Hopital's rule. For example, for $\mu_1=\mu_2$ one has
\begin{eqnarray}
\label{eq:zero-mode-deg}
\mathcal{Z}^\nu_{1,1+N_f}(\mu_b |\mu_v ,\{\mu_{sea}\})|_{\mu_1=\mu_2}
&=&
\frac{1}{2\prod_{i=1}^{N_f}(\mu_{i}^{2}-\mu_v^2)
  \prod_{j>k, j\geq 3}^{N_f}(\mu_{j}^{2}-\mu_{k}^{2})}
\nonumber\\
&&\hspace{-1.5in}
\times \det \left(
\begin{array}{ccccc}
K_\nu(\mu_{b}) & I_\nu(\mu_v) & I_\nu(\mu_{1}) & 
\mu_1^{-1}I_{\nu-1}(\mu_{1}) & \cdots\\
-\mu_{b} K_{\nu+1}(\mu_{b}) & \mu_v I_{\nu+1}(\mu_v) & \mu_{1}I_{\nu+1}(\mu_{1}) 
& I_{\nu}(\mu_{1}) & \cdots\\
\mu_{b}^2 K_{\nu+2}(\mu_{b}) & \mu_v^2I_{\nu+2}(\mu_v) & 
\mu_{1}^2I_{\nu+2}(\mu_{1}) 
& \mu_{1}I_{\nu+1}(\mu_{1}) & \cdots\\
\cdots & \cdots & \cdots & \cdots & \cdots
\end{array}
\right).
\end{eqnarray}


Partially quenched observables are computed by differentiating
Eq.~(\ref{eq:zero-mode1}) 
with respect to suitable sources and subsequently 
taking the limit $\mu_b\to \mu_v$.
For example, the zero-mode integral of Eq.~(\ref{eq:cond}) is
\begin{eqnarray}
\frac{1}{2}\langle U_{vv}+U^\dagger_{vv}\rangle_U
=\hat{\Sigma}^{\rm PQ}_\nu(\mu_v,\{\mu_{sea}\})
\equiv
-\lim_{\mu_b \to \mu_v}\frac{\partial}{\partial \mu_b} 
\ln \mathcal{Z}^\nu_{1,1+N_f}(\mu_b | \mu_v, \{\mu_{sea}\}).
\end{eqnarray}\\

To obtain the result in a $\theta=0$ (or even $\theta\neq 0$)
QCD vacuum, one could numerically sum over 
topology with the weight 
given by the partition function above;
\begin{eqnarray}
\label{eq:cond-theta}
\hat{\Sigma}_\theta^{\rm PQ}(\mu_v, \{\mu_{sea}\})
&=&\frac{\sum_\nu e^{i\theta \nu}
\hat{\Sigma}_\nu^{\rm PQ}(\mu_v, \{\mu_{sea}\})
\mathcal{Z}^\nu_{0,N_f}}{\sum_\nu e^{i\theta \nu} \mathcal{Z}^\nu_{0,N_f}},
\end{eqnarray}
or use the analytic expressions known for the $N_f=2,3$ cases 
\cite{Lenaghan:2001ur}.

\section{Doing the zero-mode integrals first: a $U(1)$ toy model}
\label{app:zero-mode-first}
\setcounter{equation}{0}

The perturbative scheme presented in this paper relies on
the operator
$$
\frac{\Sigma}{2F^2}
[\mathcal{M}^\dagger (U-1)\xi^2+\xi^2(U^\dagger-1)\mathcal{M}], 
\label{eq:U-1}
$$
being small, and more precisely of ${\cal O}(p^5)$ or smaller, for any
value of the mass $m$ as this mass is taken to zero at finite volume.
The difficulty in giving a precise counting to this term lies of
course in it being a combination of mass $m$, field $\xi(x)$ (both of
which can be assigned clear countings) {\em and} the zero mode field
$U$. It is therefore interesting that an alternative scheme exists,
which gives identical results, but which assigns a definite magnitude
to this term. This is achieved by first doing the zero
mode integral exactly, and only subsequently performing the perturbative
evaluations of the non-zero mode integrals\footnote{We are grateful to 
M. L\"{u}scher for suggesting this alternative method.}. In the
general $SU(N_f+N)$ case this becomes rather cumbersome in practice,
but it has the advantage that all terms are explicitly ordered according
to the expansion parameter $1/L$. Here we illustrate it for the case
of a simple $U(1)$ toy model at the topological sector with $\nu=0$,
where the zero-mode integrals are trivially
performed.

We thus consider the $U(1)$ ``chiral Lagrangian''\footnote{Of course,
in the real $N_f=1$ theory chiral symmetry is broken explicitly by
the anomaly, and this Lagrangian is therefore not relevant for describing the
the $N_f=1$ theory. We use it only as a simple toy model to
illustrate in a very transparent way the effect of first integrating
over the zero mode.}
\begin{equation}
{\cal L} ~=~ \frac{F^2}{4}\partial_{\mu}U(x)\partial_{\mu}U^{\dagger}(x)
- \frac{\Sigma}{2}[m^{\dagger}e^{i\theta}U(x) + {\rm h.c.}] + \cdots,
\end{equation}
where $U(x) \in U(1)$. We separate into zero modes and non-zero modes,
\begin{equation}
U(x) ~=~ Ue^{i\sqrt{2}\xi(x)/F} ~=~ e^{i\sqrt{2}\xi_0/F}e^{i\sqrt{2}\xi(x)/F},
\end{equation}
and do the analogue of fixing topology to $\nu =0$ by integrating
over $\theta$. Expanding perturbatively in the non-zero modes $\xi(x)$,
we get
\begin{eqnarray}
{\cal L} =  -\Sigma m \cos(\sqrt{2}\xi_0/F)\left(1 -
\xi(x)^2/F^2\right) + \frac{1}{2}(\partial_{\mu}\xi(x))^2 + \cdots,
\end{eqnarray}  
We now perform the zero-mode integration over $\xi_0$ exactly
with respect to the term shown to get the partition function
\begin{eqnarray}
{\cal Z} = \int [d\xi(x)] I_0\left(m\Sigma V
\left(1 -
\frac{1}{VF^2}\int d^4x \xi(x)^2\right)\right)
e^{-\frac{1}{2}\int d^4x (\partial_{\mu}\xi)^2} + \cdots,
\end{eqnarray}
up to overall irrelevant factors.

Expanding the Bessel function using $I_0'(x) = I_1(x)$, and
exponentiating the expanded term leads to the effective
partition function
\begin{eqnarray}
{\cal Z} = I_0(m\Sigma V)\int [d\xi(x)]
e^{-\int d^4x [\frac{1}{2}(\partial_{\mu}\xi)^2 
+ \frac{1}{2}\tilde{m}_{\pi}^2\xi(x)^2]} + \cdots,
\end{eqnarray}
where an effective volume-dependent pion mass is defined by
\begin{equation}
\tilde{m}_{\pi}^2 ~\equiv~ \frac{2m\Sigma}{F^2}~\frac{
I_1(m\Sigma V)}{I_0(m\Sigma V)},
\end{equation}
we note that, as expected, this effective pion mass approaches
the standard tree-level pion mass expression in the limit where
$m\Sigma V \to \infty$:
\begin{equation}
\tilde{m}_{\pi}^2 ~\to~ 2m\Sigma/F^2 = m_{\pi}^2 ~~~~~~~~
{\rm as}~~ m\Sigma V \to \infty,
\end{equation}
while in the opposite limit we find
\begin{equation}
\tilde{m}_{\pi}^2 ~\sim~ \frac{2m\Sigma}{F^2}~\frac{1}{2}m\Sigma V
= m_{\pi}^4~\frac{F^2V}{4}  ~~~~~~~~
{\rm as}~~ m\Sigma V \to 0,
\end{equation}

Clearly, in the usual $p$-regime expansion this provides us with
the standard massive propagator term, while in the chiral limit
taken at finite volume $V$, there is no infrared problem due to
the momentum sums being taken over non-zero modes only. 
When $m\Sigma V$ 
becomes of order unity we recover the usual $\epsilon$-regime expressions. 

But here we are interested in seeing how our expansion looks if
we add and subtract the standard mass term, as is done in the
main part of this paper. We therefore do a trivial rewriting,
\begin{eqnarray}
{\cal L} & = & \frac{1}{2}(\partial_{\mu}\xi(x))^2 + 
\frac{1}{2}\tilde{m}_{\pi}^2\xi(x)^2 \cr
& = &  \frac{1}{2}(\partial_{\mu}\xi(x))^2 + 
\frac{1}{2}m_{\pi}^2\xi(x)^2 + \frac{1}{2}\delta m_{\pi}^2\xi(x)^2, 
\end{eqnarray}
where
\begin{equation}
\delta m_{\pi}^2 ~\equiv~ \tilde{m}_{\pi}^2 - m_{\pi}^2,
\end{equation}

Can we treat $\frac{1}{2}\delta m_{\pi}^2\xi(x)^2$ as a perturbation? Near
the usual $p$-regime where $m\Sigma V \to \infty$ we can use the
asymptotic expansion of Bessel functions,
\begin{equation}
I_n(x) \sim \frac{e^x}{\sqrt{2\pi x}}\left(1 -
\frac{4n^2 - 1}{8x} + \cdots \right),
\end{equation}
to see that
\begin{equation}
\delta m_{\pi}^2 = - \frac{1}{F^2V} + \cdots,
\end{equation}
This is of ${\cal O}(p^4)$ as expected in this $U(1)$ theory.
When instead $m\Sigma V$ is of order unity, we get
\begin{equation}
\delta m_{\pi}^2 \sim - m_{\pi}^2,
\end{equation}
which is also of ${\cal O}(p^4)$. The point here is that we know
the full analytical expression
\begin{equation}
\delta m_{\pi}^2 ~=~ \frac{2m\Sigma}{F^2}\left[\frac{
I_1(m\Sigma V)}{I_0(m\Sigma V)} - 1\right],
\end{equation}
for all values of $m$ and $V$, and it is easy to check that
this function is of ${\cal O}(p^4)$ everywhere. The term
$\delta m_{\pi}^2\xi(x)^2$ is thus explicitly found to be
of NLO and we can treat it perturbatively.

Of course, as already discussed in Section 5, away from the
conventional $p$-regime expansion, separating out 
$\delta m_{\pi}^2\xi(x)^2$ and treating it perturbatively
amounts to a re-ordering and partial resummation of terms in 
this expansion. This is because the propagator is taken to be
the conventional massive one even when $m_{\pi}^2$ is no longer
of ${\cal O}(p^2)$, but {\em smaller}. The difference between
a calculation based on this propagator and one where the
propagator has been expanded in $m_{\pi}^2$ up to the needed
order is always of yet higher order, and thus only illustrates the
inherent uncertainty in any fixed-order perturbative expansion.

\section{Numerical evaluation of $\bar{\Delta}(0,M^2)$ and 
$\partial_{M^2}\bar{\Delta}(0,M^2)$}
\label{app:g1g2}

The definition of $\bar{g}_1(M^2)$ in Eq.~(\ref{eq:g1})
and $\bar{g}_2(M^2)=-\partial_{M^2}\bar{g}_1(M^2)$,
requires an infinite sum over the 4-vector 
$a_\mu = n_\mu L_\mu$ where $L_i = L\; (i=1,2,3)$ and $L_4=T$.
In this appendix, we will suggest how to 
evaluate numerically $\bar{g}_1(M^2)$ and $\bar{g}_2(M^2)$.\\

For $ML>1$, the expansion 
\cite{BesselfiniteV},
\begin{eqnarray}
\label{eq:g1p}
\bar{g}_1(M^2) = \sum_{a\neq 0}\int 
\frac{d^4 q}{(2\pi)^4}\frac{e^{-iqa}}{q^2+M^2}-\frac{1}{M^2V}
=\sum_{a\neq 0} \frac{\sqrt{M^2}}{4\pi^2|a|}K_1(\sqrt{M^2}|a|)-\frac{1}{M^2V},
\end{eqnarray}
in terms of modified Bessel functions is useful,
while for $ML<1$, the polynomial expression
\begin{eqnarray}
\label{eq:g1e}
\bar{g}_1(M^2) = -\frac{M^2}{16\pi^2}\ln (M^2 V^{1/2})
-\sum^{\infty}_{n=1}\frac{\beta_n}{(n-1)!} M^{2(n-1)}V^{(n-2)/2},
\end{eqnarray}
using the shape coefficients $\beta_n$'s 
\cite{Hasenfratz:1989pk} is appropriate. Here
\begin{eqnarray}
\beta_n &\equiv  & \left(\frac{-1}{4\pi}\right)^n \left(
\alpha_n+\frac{2}{n(n-2)}\right)\;\;\; (n\neq 2),\;\;\;\;\;
\beta_2 \equiv \frac{\alpha_2 - \ln 4\pi + \gamma - 3/2}{16\pi^2},\\
\alpha_n &\equiv & 
\int_0^1 dt\left\{t^{n-3}\left(S\left(\frac{L^2}{V^{1/2}t}\right)^3
S\left(\frac{T^2}{V^{1/2}t}\right)-1\right)
\right.\nonumber\\&&\left.\hspace{1in}
+
t^{-n-1}\left(S\left(\frac{V^{1/2}}{L^2t}\right)^3
S\left(\frac{V^{1/2}}{T^2t}\right)-1\right)\right\},\\
S(x) &\equiv & \sum^\infty_{k=-\infty} \exp (-\pi k^2 x),
\end{eqnarray}
where $\gamma\sim 0.577215665$ is Euler's constant 
and the summation in $S(x)$
is typically well approximated by a truncation
to $|k| \le 20$.

Since the modified Bessel function $K_\nu(x)$ is well-defined
for the complex value of $x$, the both expressions above
can be easily extended for the complex arguments using
$(m+\sqrt{m_v^2})^{1/2}|_{m_v=i\lambda}=(m^2+\lambda^2)^{1/4}
e^{i\arctan(\lambda/m)/2}$,
and the prescription given in Eq.~(\ref{eq:analcont2}).

As shown in Fig.~\ref{fig:g1g2},
ignoring the contribution from
$|n_\mu| > 20$ in the summation in Eq.~(\ref{eq:g1p}),
and $n > 5$ in the summation in Eq.~(\ref{eq:g1e}),
both formulas agree
in a rather long interval around $ML=1$.
We also find an excellent agreement between these
two expansions for both ${\rm Re}\bar{g}_1(M^2)$ 
and ${\rm Re}\bar{g}_2(M^2)$ at
imaginary values of $M^2$, using the same truncation.
It is in any case trivial to increase the accuracy by including more terms. 

To summarize, we have used
\begin{eqnarray}
\label{eq:g1nume}
\bar{g}_1(M^2) = \left\{
\begin{array}{lc}
\bar{g}^p_1(M^2)\equiv \sum_{a
\neq 0}^{|n_i| \leq 20} 
\frac{\sqrt{M^2}}{4\pi^2|a|}K_1(\sqrt{M^2}|a|)-\frac{1}{M^2V}& (|M|L>1)\\\\
\bar{g}^\epsilon_1(M^2)\equiv-\frac{M^2}{16\pi^2}\ln (M^2 V^{1/2})
-\sum^{5}_{n=1}\frac{\beta_n}{(n-1)!} M^{2(n-1)}V^{(n-2)/2} &(|M|L\leq 1)
\end{array}
\right. ,\nonumber\\
\end{eqnarray}
and similarly for their derivatives $\bar{g}_2$.
We plot 
${Re}\bar{g}_1(M^2_0+i M^2)$ and ${Re}\bar{g}_2(M^2_0+i M^2)$
for various choices of fixed $M_0$ in Fig. \ref{fig:Reg1Reg2}.

Specifying the renormalization scale $\mu_{sub}=0.77$ GeV of
the low energy constants $L_i$'s, 
and the above prescription for $\bar{g}_1$ and $\bar{g}_2$,
\begin{eqnarray}
\bar{\Delta}(0,M^2) &=& \frac{M^2}{16\pi^2}\ln \frac{M^2}{\mu_{sub}^2}
+\bar{g}_1(M^2),
\\
\partial_{M^2}\bar{\Delta}(0,M^2)&=&
\frac{1}{16\pi^2}\left(\ln \frac{M^2}{\mu_{sub}^2}+1\right)
-\bar{g}_2(M^2),
\end{eqnarray}
can then be evaluated numerically.
We note that in the small mass region the logarithmic contribution
$\ln M^2$ is canceled by a similar one in $\bar{g}_i$ and that
both $\bar{\Delta}(0,M^2)$ and   $\partial_{M^2}\bar{\Delta}(0,M^2)$
have no IR divergences in the limit $M^2\to 0$.

\FIGURE[t]{
\epsfig{file=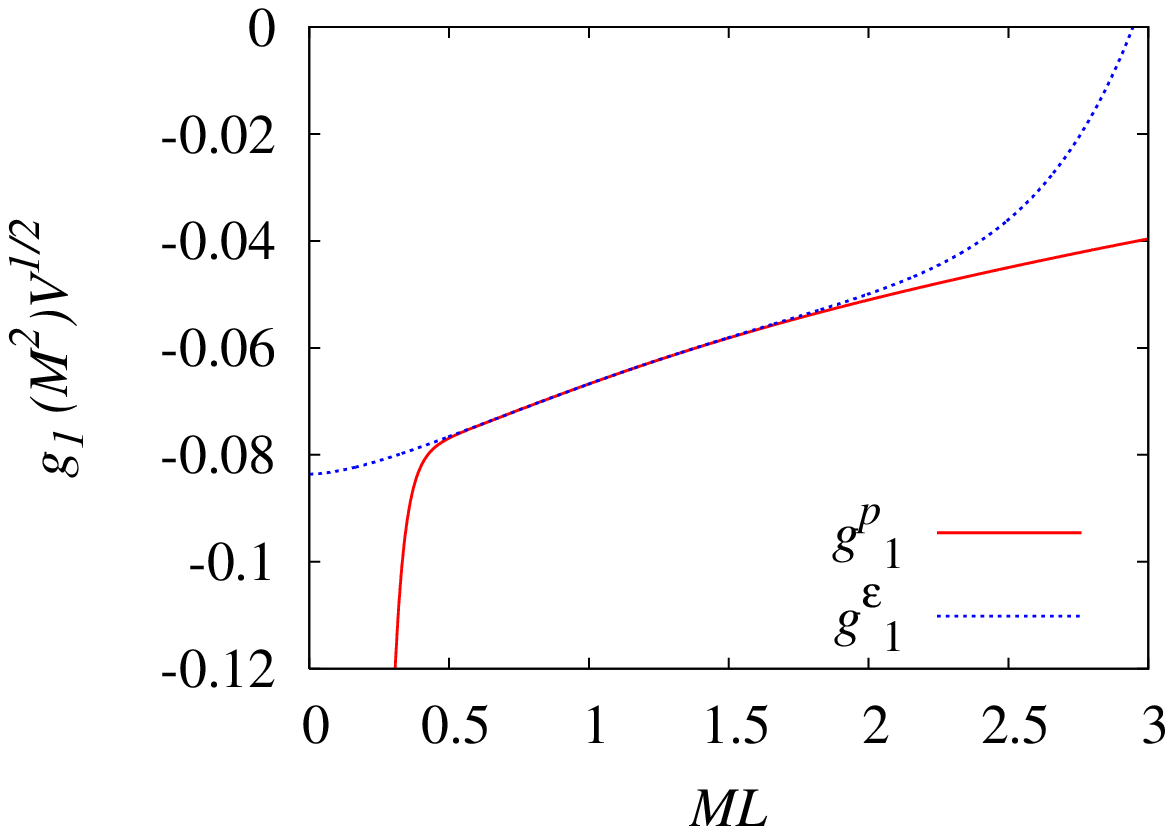,width=7.4cm}
\epsfig{file=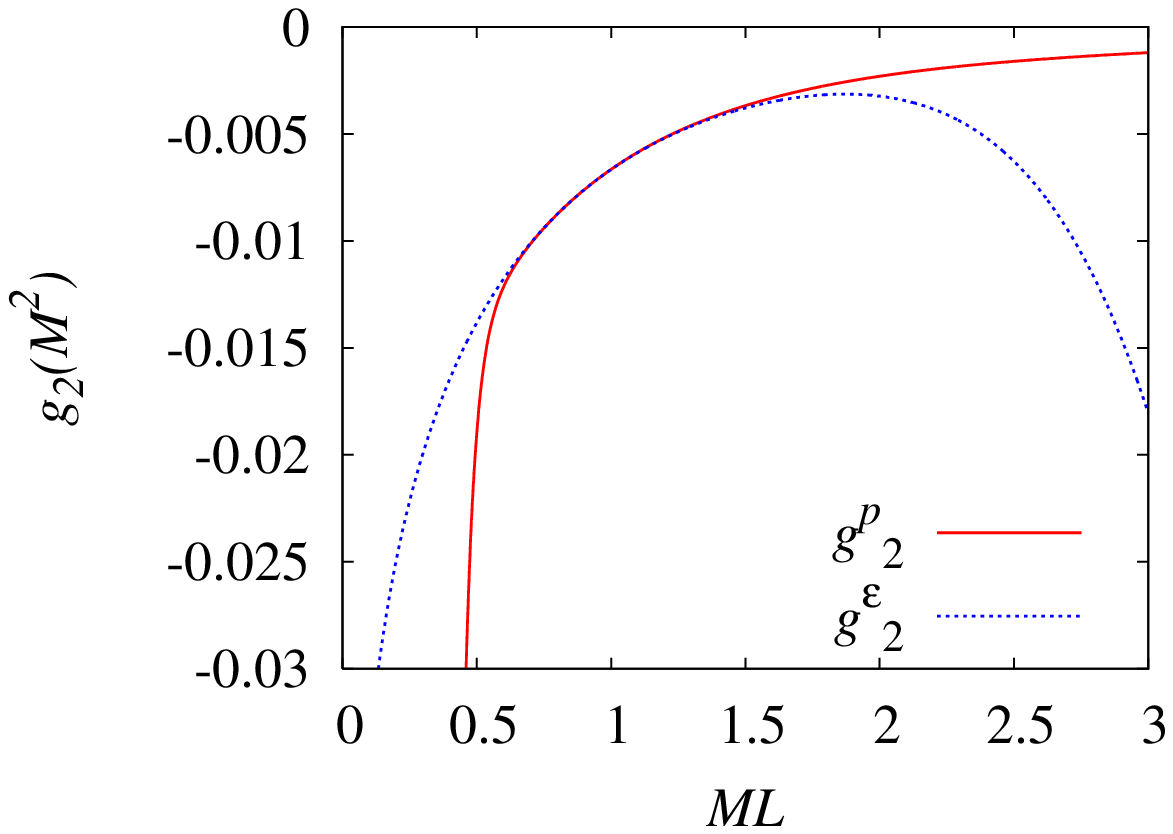,width=7.4cm}
  \caption{
    The plot of $\bar{g}_1(M^2)V^{1/2}$ (left part)
    and $\bar{g}_2(M^2)$ (right part) for $T=2L$.
    The truncated sums in Eq.~(\ref{eq:g1nume})
    for $\bar{g}^p_{1,2}(M^2)$ (solid line) and 
    $\bar{g}^\epsilon_{1,2}(M^2)$
    (dotted) agree well around $ML\sim 1$.
    Note that $\bar{g}_1(M^2)V^{1/2}$ converges to 
    $-\beta_1=-0.08360$ in the chiral limit.
    Note also that $|\bar{g}_1(M^2)|$ attains
    its maximum at $M=0$.
}
  \label{fig:g1g2}
}
%
\FIGURE[t]{
\epsfig{file=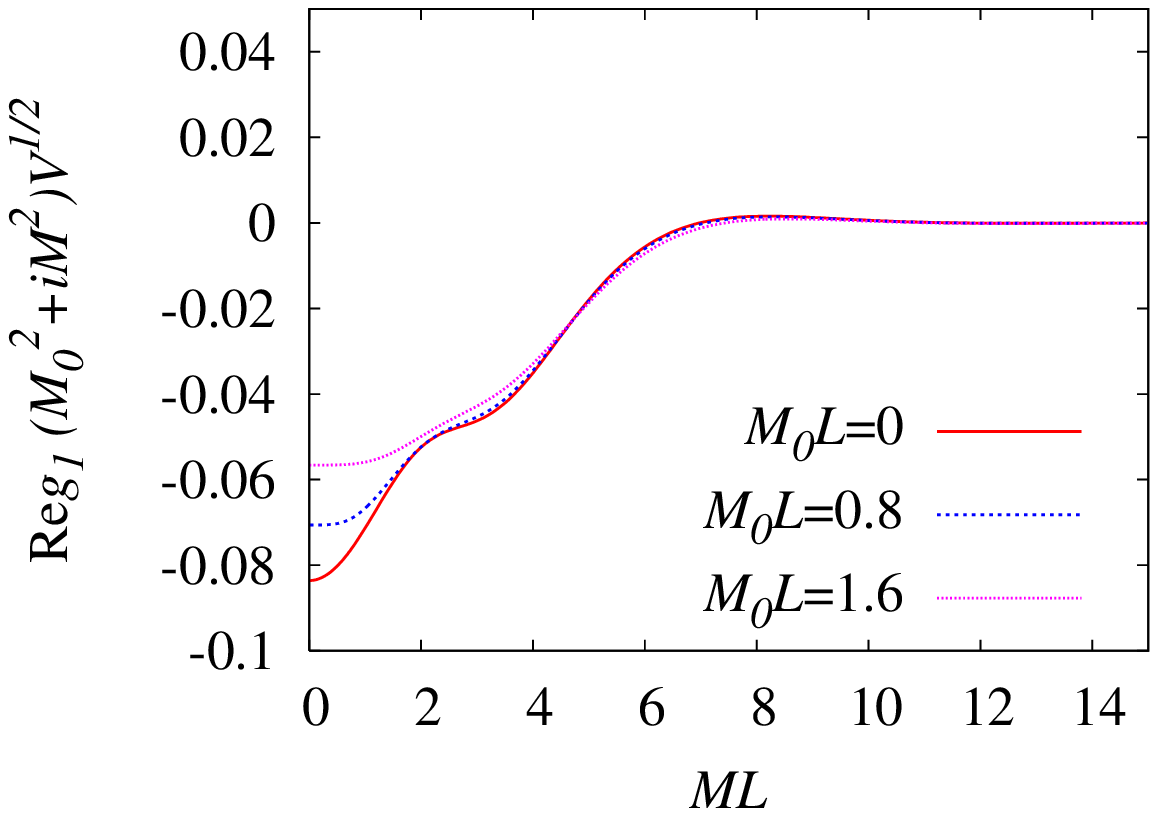,width=7.4cm}
\epsfig{file=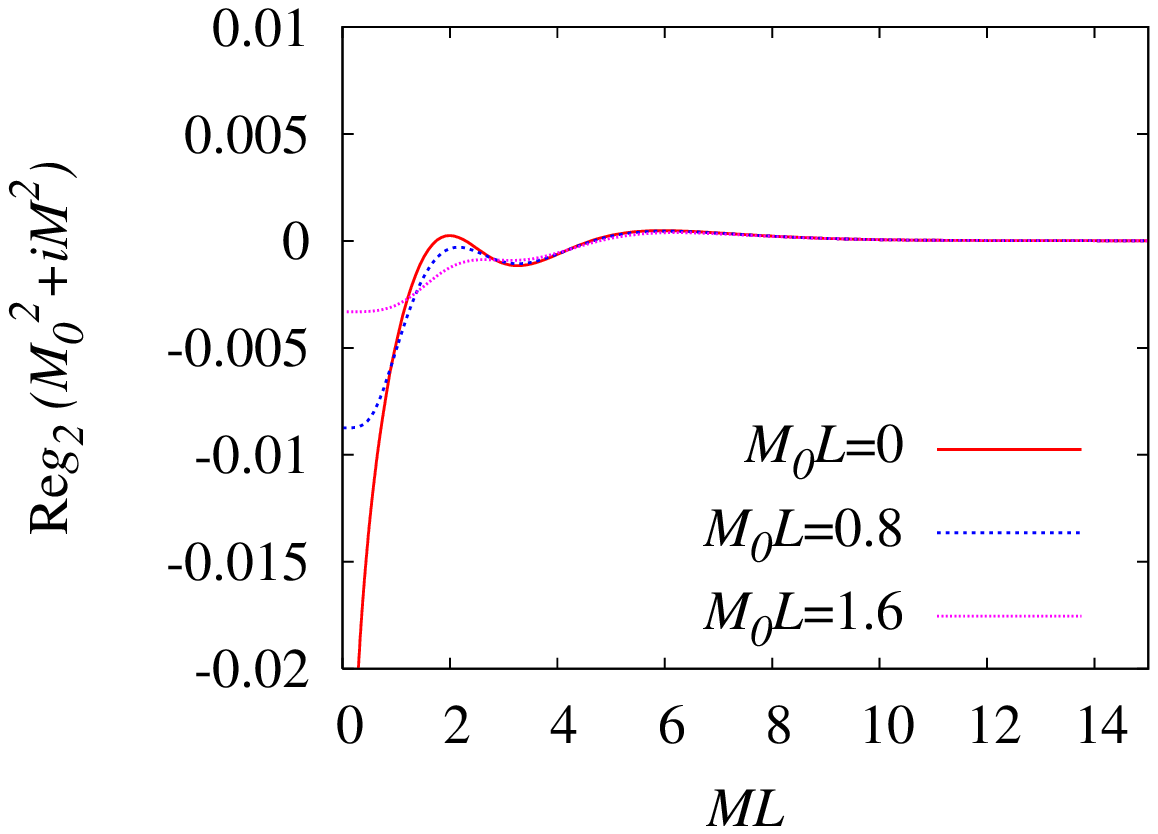,width=7.4cm}
  \caption{
    Plot of ${\rm Re}\bar{g}_1(M^2_0+i M^2)V^{1/2}$ (left part)
    and ${\rm Re}\bar{g}_2(M^2_0+iM^2)$ (right part) for $T=2L$.
    for three fixed values of $M_0L=0$ (solid), 0.05 (dotted)
    and 0.1 (small dotted).
}
  \label{fig:Reg1Reg2}
}

\section{Useful properties of $\bar{G}(x, M^2,M^2)$
}
\label{app:G-property}

We also need the 1-loop correction from off-diagonals part of
$\xi$, $\bar{G}(0,M^2,M^2)$.
It can, in principle, be expressed in terms of $\bar{\Delta}(0,M^2)$.
In this appendix, we discuss the UV divergence of
$\bar{G}(0,M^2,M^2)$ and rewrite it using   
$\bar{\Delta}(0,M^2)$ and $\partial_{M^2}\bar{\Delta}(0,M^2)$.
Explicit examples for the degenerate case 
and non-degenerate $N_f=2+1$ flavor theory will be given.\\

First let us consider the UV-divergent part of $\bar{\Delta}(0,M^2)$,
\begin{eqnarray}
\bar{\Delta}(0,M^2) &=& \frac{1}{V}\sum_{p\neq 0}\frac{1}{p^2+M^2}
\nonumber\\&=&
\frac{1}{V}\sum_{p\neq 0}\left(\frac{1}{p^2}-
\frac{M^2}{p^4}\right)+{\cal O}(M^4) .
\end{eqnarray}
Here the first term is quadratically divergent, 
the second term produces a logarithmic divergence, 
and the remaining ${\cal O}(M^4)$ terms are UV finite.
As is well-known. the quadratic divergence is absent when we employ 
dimensional regularization.\\

By expanding in the mass, the UV-divergent part of 
$\bar{G}(0,M^2,M^2)$ can be written
\begin{eqnarray}
\label{eq:Gdiv}
\bar{G}(0,M^2,M^2) &=& \frac{1}{V}\sum_{p\neq 0}
\frac{1}{(p^2+M^2)^2\left(\sum_f^{N_f}\frac{1}{p^2+M^2_{ff}}\right)}
\nonumber\\&=&
\frac{1}{N_fV}\sum_{p\neq 0}\left(\frac{1}{p^2}-
\frac{2M^2}{p^4}+\frac{1}{N_f}\sum^{N_f}_f\frac{M^2_{ff}}{p^4}
+{\cal O}(M^4)\right)\nonumber\\
&=&\frac{2}{N_f}\bar{\Delta}(0,M^2)
-\frac{1}{N_f^2}\sum^{N_f}_f\bar{\Delta}(0,M^2_{ff})+{\cal O}(M^4)
\nonumber\\
&=& \left(\frac{2}{N_f}M^2-\frac{1}{N_f^2}\sum^{N_f}_f M^2_{ff}\right)\frac{c_1}{16\pi^2}+\cdots,
\end{eqnarray}
where the logarithmic divergence of the last line is canceled by
a renormalization of $L_i$'s as seen in Section \ref{sec:condensate}.\\

Although Eq.~(\ref{eq:Gdiv}) shows that $\bar{G}(0,M^2,M^2)$
inherits the UV properties of $\bar{\Delta}(0,M^2)$,
the expansion is not useful when we want to see the finite part of 
$\bar{G}(0,M^2,M^2)$, since the omitted ${\cal O}(M^4)$ terms 
are the at the same order as the former two terms.
In order to obtain a convenient rewriting, let us define a function
\begin{eqnarray}
\label{eq:f(t)}
f(t) \equiv \frac{1}{N_f}\sum^k_i \frac{n_i}{t-M^2_{ii}},
\end{eqnarray}
where $k$ denotes the number of different quark masses and
$n_i\geq 1$ is the degeneracy of the $i$-th mass
satisfying $\sum^k_in_i=N_f$.
Here we have ordered the masses $M^2_{ii}< M^2_{i+1\;i+1}$ for any $i$.
Noting $f(t)$ is a monotonically decreasing function,
\begin{eqnarray}
\frac{d}{dt}f(t) = -\frac{1}{N_f}
\sum^k_i \frac{n_i}{(t-M^2_{ii})^2}<0,
\end{eqnarray}
and
\begin{eqnarray}
\lim_{\epsilon \to 0}f(M_{ii}^2+\epsilon)&=&\infty,\;\;\;
\lim_{\epsilon \to 0}f(M_{i+1\;i+1}^2-\epsilon)=-\infty,\\
f(t) &\neq& 0,\;\;\; \mbox{for}\;\;\; t<M^2_{11}, M^2_{kk}<t,
\end{eqnarray}
one can show that an equation $f(t)=0$ has $k-1$ different
solutions (we denote them by $t=\bar{M}^2_{ii}$), each of them satisfying
\begin{eqnarray}
M^2_{ii}<\bar{M}^2_{ii}<M^2_{i+1\;i+1},\;\;\;(1\leq i \leq k-1).
\end{eqnarray}
We illustrate this in the plot Fig.~\ref{fig:f(t)}.

Hence, 
\begin{eqnarray}
-f(-p^2)=\frac{\prod_i^{k-1}(p^2+\bar{M}^2_{ii})}
{\prod_j^{k}(p^2+M^2_{jj})},
\end{eqnarray}
and $\bar{G}(0,M^2,M^2)$ can thus alternatively be expressed as
\begin{eqnarray}
\bar{G}(0,M^2,M^2)&=&\frac{1}{N_fV}\sum_{p\neq 0}
\frac{\prod_j^{k}(p^2+M^2_{jj})}
{(p^2+M^2)^2\prod_i^{k-1}(p^2+\bar{M}^2_{ii})}
\nonumber\\
&=& \frac{1}{N_f}\left[ \sum^{k-1}_i A_i \bar{\Delta}(0,\bar{M}^2_{ii})
+B \bar{\Delta}(0,M^2) +
C\;\partial_{M^2}\bar{\Delta}(0,M^2)  \right],
\end{eqnarray}
where the coefficients $A_i$'s, $B$ and $C$ are
given by the residue of 
\begin{eqnarray}
f_2(t)=\frac{\prod_j^{k}(-t+M^2_{jj})}
{(-t+M^2)^2\prod_i^{k-1}(-t+\bar{M}^2_{ii})},
\end{eqnarray}
(or $-(-t+M^2)f_2(t)$ for $C$), at each pole.
Note that $C=0$ when $M^2$ is equal to
any of the physical masses.\\

Noting that both $\bar{\Delta}(0,M^2)$ and
$\partial_{M^2}\bar{\Delta}(0,M^2)$ are infra-red finite
in the limit $M^2 \to 0$,
\begin{eqnarray}
\bar{\Delta}(0,M^2)|_{M^2\to 0}&=& -\frac{\beta_1}{\sqrt{V}},\\
\partial_{M^2}\bar{\Delta}(0,M^2)|_{M^2\to 0}
&=&-\frac{1}{16\pi^2}\ln V^{1/2}-\beta_2 +\frac{c_1}{16\pi^2},
\end{eqnarray}
the chiral limit of $\bar{G}(0,M^2,M^2)$ is given by
\begin{eqnarray}
\bar{G}(0,0,0)&=&
\frac{1}{N_f}\left[ \sum^{k-1}_i A_i|_{M^2=0} 
\bar{\Delta}(0,\bar{M}^2_{ii})
-B|_{M^2=0}   \frac{\beta_1}{\sqrt{V}}
\right.\nonumber\\&&\hspace{1in}\left.
-C|_{M^2=0} \left(\frac{1}{16\pi^2}\ln V^{1/2}+\beta_2
-\frac{c_1}{16\pi^2}\right)
\right],
\end{eqnarray}
where UV divergence $c_1$ is absorbed 
into the renormalization of $L_6$.

Here we give some useful examples.
For the fully degenerate case, $i.e.$, equal masses 
$M^2_{ii} = M^2_{sea}$ for all $i$, 
the above expression for $\bar{G}$ is greatly simplified,
\begin{eqnarray}
\bar{G}(0,M^2,M^2)
&=& \frac{1}{N_f}\left[ 
\bar{\Delta}(0,M^2) +
(M^2-M_{sea}^2)\partial_{M^2}\bar{\Delta}(0,M^2)  \right] ,
\end{eqnarray}
in agreement with the result presented in ref. \cite{Osborn:1998qb}.

For an $N_f=2+1$ flavor theory, 
the equation $f(t)=0$ is easily solved 
and one obtains
\begin{eqnarray}
\label{eq:Gbar2+1}
\bar{G}(0,M^2,M^2)
&=& \frac{1}{3}\left[
A  \bar{\Delta}\left(0,\frac{M^2_{ud}+2M^2_{ss}}{3}\right)
+B \bar{\Delta}(0,M^2) +
C\;\partial_{M^2}\bar{\Delta}(0,M^2)  \right],
\nonumber\\
\end{eqnarray}
where $M^2_{ud}=2m_u\Sigma/F^2=2m_d\Sigma/F^2$, 
$M^2_{ss}=2m_s\Sigma/F^2$ and the coefficients are given by
\begin{eqnarray}
\label{eq:ABC2+1}
A &=& -\frac{2(M^2_{ud}-M^2_{ss})^2}
{(3M^2-M^2_{ud}-2M^2_{ss})^2},\;\;\;
B = 1 + \frac{2(M_{ud}^2-M_{ss}^2)^2}{(3M^2-M^2_{ud}-2M^2_{ss})^2},
\nonumber\\
C&=&\frac{3(M^2-M^2_{ud})(M^2-M^2_{ss})}{3M^2-M^2_{ud}-2M^2_{ss}}.
\end{eqnarray}

\FIGURE[t]{
\epsfig{file=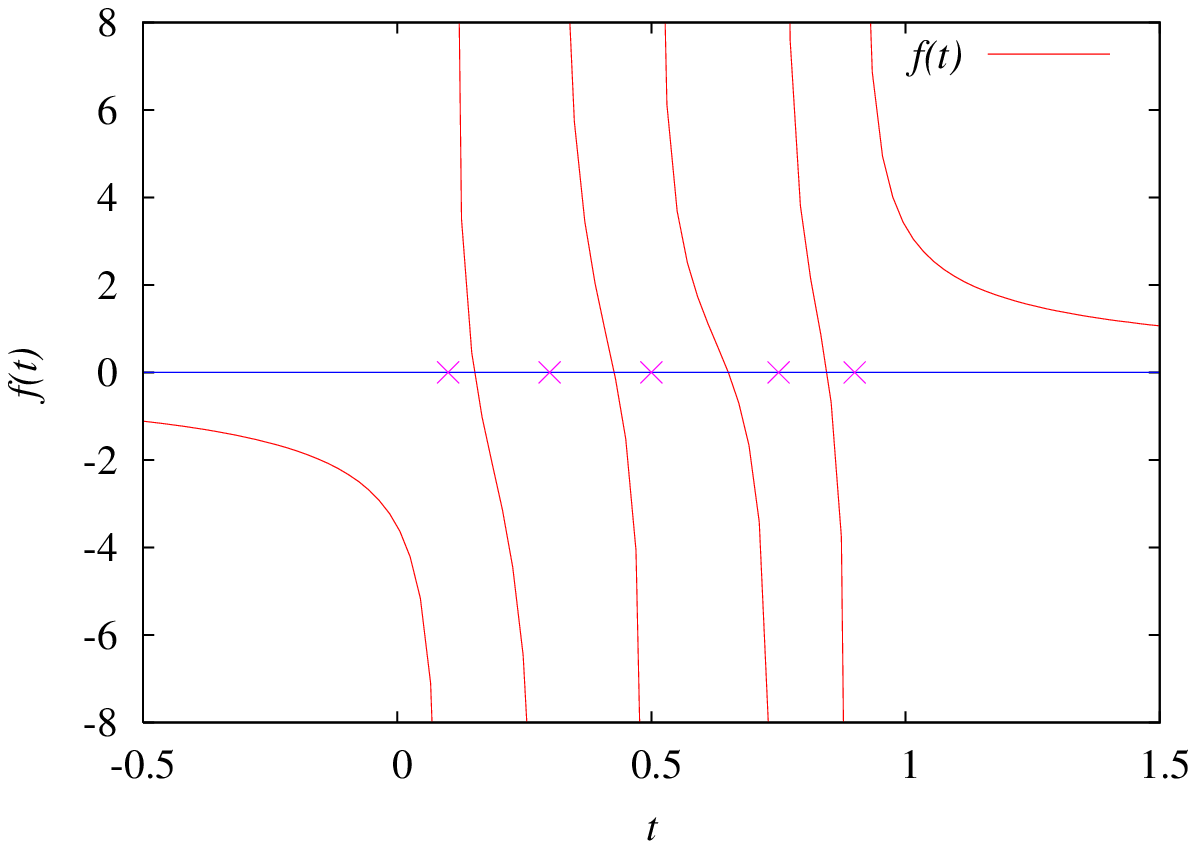, width=9cm}
  \caption{
    Example of $f(t)$ in Eq.~(\ref{eq:f(t)}), where we have $N_f=6$, $k=5$,
    $n_i=1 (i\neq 2)$, $n_2=2$, and $\{M^2_{ii}\}=\{0.1,0.3,0.5,0.75,0.9\}$.
    One can see $f(t)=0$ has $k-1=4$ solutions between the poles 
    (plotted by the crosses).
  }
  \label{fig:f(t)}
}

\end{document}